\documentclass[aps,prb,superscriptaddress,showpacs,floatfix]{revtex4-1}
\usepackage{graphicx,amsmath,amssymb,xspace,epsfig,float,multirow,subfigure,tabularx}

\pdfoutput=1

\begin{document}

\title{Magnetic properties of Type I and II Weyl Semi-metals in
Superconducting state.}
\author{Baruch Rosenstein}
\email{baruchro@hotmail.com}
\affiliation{Electrophysics Department, National Chiao Tung University, Hsinchu 30050,
\textit{Taiwan, R. O. C}}
\author{B.Ya. Shapiro}
\email{shapib@mail.biu.ac.il}
\affiliation{Physics Department, Bar-Ilan University, 52900 Ramat-Gan, Israel}
\author{Dingping Li}
\email{lidp@pku.edu.cn}
\affiliation{School of Physics, Peking University, Beijing 100871, \textit{China}}
\affiliation{Collaborative Innovation Center of Quantum Matter, Beijing, China}
\author{I. Shapiro}
\email{yairaliza@gmail.com}
\affiliation{Physics Department, Bar-Ilan University, 52900 Ramat-Gan, Israel}
\date{\today }

\begin{abstract}
Superconductivity was observed in certain range of pressure and chemical
composition in Weyl semi-metals of both the type I and type II (when the
Dirac cone tilt parameter $\kappa >1$). Magnetic properties of these
superconductors are studied on the basis of microscopic phonon mediated
pairing model. The Ginzburg - Landau effective theory for the order
parameter is derived using Gorkov approach and used to determine anisotropic
coherence length, the penetration depth determining the Abrikosov parameter
for a layered material and applied to recent extensive experiments on $%
MoTe_{2}$. It is found that superconductivity is of second kind near the
topological transition at $\kappa =1$. For a larger tilt parameter
superconductivity becomes first kind. For $\kappa <1$ the Abrikosov
parameter\ also tends to be reduced, often crossing over to the first kind.
For the superconductors of the second kind the dependence of critical fields
$H_{c2}$ and $H_{c1}$ on the tilt parameter $\kappa $ (governed by pressure)
is compared with the experiments. Strength of thermal fluctuations is
estimated and its is found that they are strong enough to cause Abrikosov
vortex lattice melting near $H_{c2}$. The melting line is calculated and is
consistent with experiments provided the fluctuations are three dimensional
in the type I phase (large pressure) and two dimensional in the type II
phase (small pressure).
\end{abstract}

\maketitle

\section{Introduction}

Dispersion relation near Fermi surface in recently synthesized two and three
dimensional Weyl (Dirac) semi-metals\cite{Weng,MoTeearly,ZrTe} is
qualitatively distinct from conventional metals, semi - metals or
semiconductors, in which all the bands are parabolic. In type I Weyl
semi-metals (WSM), the band inversion results in Weyl points in low-energy
excitations being anisotropic massless "relativistic" fermions. They exhibit
several remarkable properties like the chiral magnetic effect\cite{chiral}
related to the chiral anomaly in particle physics. More recently, type-II
WSMs, layered transition-metal dichalcogenides, were discovered\cite%
{Soluyanov}. Here, the Weyl cone exhibits such a strong tilt, so that they
can be characterized by a nearly flat band at Fermi surface. The type-II WSM
also exhibit exotic properties different from the type-I ones, such
anti-chiral effect of the chiral Landau level,\cite{Yu} and novel quantum
oscillations \cite{Brien}.

Graphene is a prime example of the type I WSM, while materials, like layered
organic compound $\alpha -(BEDT-TTF)_{2}I_{3}$, were long suspected\cite%
{Goerbig1} to be a 2D type-II Dirac fermion. Several materials were observed
to undergo the I to II transition while doping or pressure is changed\cite%
{toptransition}. Theoretically physics of the topological (Lifshitz) phase
transitions between the type I to type II Weyl semi-metals were considered
in the context of superfluid phase\cite{Volovik} A of $He_{3}$, layered
organic materials in 2D\cite{Goerbig2} and 3D Weyl semi-metals\cite{Ye}. The
pressure modifies the spin orbit coupling that in turn determines the
topology of the Fermi surface of these novel materials \cite{Sun}.

Many Weyl materials are known to be superconducting. A detailed study of
superconductivity in WSM under hydrostatic pressure revealed a curious
dependence of critical temperature of the superconducting transition on
pressure. The critical temperature $T_{c\text{ }}$ in some of these systems
like $HfTe_{5}$ show\cite{HfTe} a sharp maximum as a function of pressure.
This contrasts with generally smooth dependence on pressure in other
superconductors (not suspected to be Weyl materials) like a high $T_{c}$
cuprate\cite{YBCO} $YBCO$. Since superconductivity is especially affected by
the type I to II topological transition, it might serve as such an indicator%
\cite{Zyuzin,Rosenstein17}.

Various mechanisms of superconductivity in WSM turned superconductors have
been considered theoretically \cite{DasSarma,FuBerg,frontiers}, however
evidence point towards the conventional phonon mediated one. If the Fermi
level is not situated too close to the Dirac point, the BCS type pairing
occurs, otherwise a more delicate formalism should be employed\cite%
{Shapiro14}. A theory predicted possibility of superconductivity in the type
II Weyl semimetals was developed recently in the framework of Eliashberg
model \cite{Zyuzin,Rosenstein17}.

In the present paper we extend the study of superconductivity in Weyl
semimetals of both types to magnetic properties and thermal fluctuations.
The phenomenological Ginzburg-Landau theory for superconducting WSM of the
arbitrary type is microscopically derived and used to establish magnetic
phase diagram. In particular the Abrikosov parameter used to distinguish
between the superconductivity of the first from the second kind is
determined. It turns out that superconductivity is of second kind near the
critical value of the tilt parameter $\kappa =1$, marking the topological
transition, but becomes first kind away from it on both the type I and type
II sides. The critical fields, coherence lengths magnetic penetration depths
and the Ginzburg number characterizing the strength of fluctuations are
found. In the strongly layered material like\cite{Tamai} $MoTe_{2}$ the
fluctuations are strong enough to qualitatively affect the Abrikosov vortex
phase diagram: the lattice "melts" into the vortex liquid \cite{MoTe2melting}%
. This is reminiscent of a well known (possibly non - Weyl semi-metal)
layered dichalcogenides superconductor $NbSe_{2}$ that is perhaps the only
low $T_{c}$ material with fluctuations strong enough to exhibit vortex
lattice melting\cite{NbSe2}. The Ginzburg number for these single crystals
is of order of $Gi=10^{-4}$ with similar $T_{c}$ and upper critical field $%
H_{c2}\left( 0\right) $ of several $Tesla$.

The focus generally is on the dependence of the properties in the cone tilt
parameter $\kappa $ and consequently on the transition from Type-I to
type-II WSM variations. This is experimentally measured in experiments on
the pressure (determining $\kappa $) dependence of WSM superconductors.
These days there are already quite a variety of WSM turned superconductors
and it is impossible to model all of them in a single paper. Therefore one
of the best studied material, $MoTe_{2}$ is chosen as a representative
example. A major reason is that magnetic properties of this superconductor
were investigated in a wide range pressures\cite{MoTe2melting} from ambient
to $30GPa$ (controlling the tilt parameter $\kappa $ of the WSM, see below).
An additional advantage of this choice is that the strongly layered material
$MoTe_{2}$ in many aspects behaves as a simpler two dimensional WSM (weak
van der Waals coupling between the layers is easily accounted for).

The paper is organized as follows. The next section contains the formulation
of a sufficiently general the phonon mediated BCS - like model of
anisotropic type I and II WSM. Gor'kov equations are written with details
relegated to appendices. The section III is devoted to derivation from the
Gor'kov equations in the inhomogeneous case of the coefficients of the
Ginzburg - Landau equations including the gradient term. Magnetic properties
are derived from the GL model in section IV, while thermal fluctuations are
subject of section V. In particular vortex lattice melting line is
considered. Section VI contains conclusions and discussion of the
experimental data on $MoTe_{2}$.

\section{Pairing in Weyl semimetal.}

\subsection{The model}

Considering layered WSM as alternating superconducting 2D layers separated
by dielectric streaks. We assume that a 3D electrons with strongly
anisotropic dispersion relation are paired inside the 2D layers only. We
start to study the effect of the topological transition on superconductivity
using the simplest possible model of a single 2D WSM layer with just two
sublattices denoted by $\alpha =1,2$ and expand this model to real 3D
layered system. The band structure near the Fermi level of a 2D Weyl
semi-metal is well captured by the non-interacting massless Weyl Hamiltonian
with the Fermi velocity $v$ (assumed to be isotropic in the $x-y$ plane) and
conventional parabolic term on $z-$direction \cite{Rosenstein17}:

\begin{eqnarray}
K &=&\int_{\mathbf{r}}\psi _{\alpha }^{s+}\left( \mathbf{r}\right) \widehat{K%
}_{\alpha \beta }\psi _{\beta }^{s}\left( \mathbf{r}\right) \text{\ \ \ \ }
\label{eq1} \\
\text{\ \ }\widehat{K}_{\gamma \delta } &=&-i\hbar v\nabla ^{i}\sigma
_{\gamma \delta }^{i}+\left( -i\hbar w_{i}\nabla ^{i}-\mu +\frac{p_{z}^{2}}{%
2m_{z}}\right) \delta _{\gamma \delta }\text{.}  \notag
\end{eqnarray}%
Here $\mu $ is the chemical potential, $p_{z}=-i\hbar \nabla _{z}$ , $\sigma
$ are Pauli matrices in the sublattice space and $s$ is spin projection. The
velocity vector $\mathbf{w}$ defines the tilt of the (otherwise isotropic)
cone. (We use below the dimensionless ratio $\kappa =w/v$ as tilt parameter
describing cone axis projection in $x$ direction). The graphene - like
dispersion relation for $\mathbf{w}=0$ represents the type I Weyl
semi-metal, while for the velocity $\left\vert \mathbf{w}\right\vert =w$
exceeding $v$, the material becomes a type II Weyl semi - metal.

Generally there are a number of pairs of points (Weyl cones) constituting
the Fermi "surface" of such a material at chemical potential $\mu =0$. We
restrict ourself to the case of just one left handed and one right handed
Dirac points, typically but not always separated in the Brillouin zone.
Generalization to include the opposite chirality and several "cones" is
straightforward. We assume that different valleys are paired independently
and drop the valley indices (multiplying the density of states by $2N_{f}$).

The effective electron-electron attraction due to the electron - phonon
attraction opposed by Coulomb repulsion (pseudopotential) mechanism creates
pairing below $T_{c}$. Further we assume the singlet $s$-channel interaction
with essentially local interaction,
\begin{equation}
V=\frac{g^{2}}{2}\ \int d\mathbf{r}\text{ }\psi _{\alpha }^{+\uparrow
}\left( \mathbf{r}\right) \psi _{\beta }^{\downarrow +}\left( \mathbf{r}%
\right) \psi _{\beta }^{\uparrow }\left( \mathbf{r}\right) \psi _{\alpha
}^{\downarrow }\left( \mathbf{r}\right) \text{,}  \label{int}
\end{equation}%
where the coupling $g^{2}$ is zero between the layers. As usual the retarded
interaction has a cutoff frequency $\Omega $, so that it is active in an
energy shell of width $2\hbar \Omega $ around the Fermi level \cite%
{Abrikosov}. For the phonon mechanism it is the Debye frequency. We first
remind\cite{Rosenstein17}, the Gorkov equations and then derive from them
the phenomenological GL equations that allow to obtain the basic magnetic
response of the superconductors.

\subsection{Green Functions and Gor'kov equations}

Finite temperature properties of the condensate are described at temperature
$T$ by the normal and the anomalous Matsubara Greens functions\cite%
{Abrikosov} (GF),%
\begin{eqnarray}
G_{\alpha \beta }^{ts}\left( \mathbf{r}\tau ,\mathbf{r}^{\prime }\tau
^{\prime }\right) &=&-\left \langle T_{\tau }\psi _{\alpha }^{t}\left(
\mathbf{r}\tau \right) \psi _{\beta }^{s+}\left( \mathbf{r}^{\prime }\tau
^{\prime }\right) \right \rangle =\delta ^{ts}g_{\alpha \beta }\left(
\mathbf{r-r}^{\prime },\tau -\tau ^{\prime }\right) ;  \label{green} \\
F_{\alpha \beta }^{ts}\left( \mathbf{r}\tau ,\mathbf{r}^{\prime }\tau
^{\prime }\right) &=&\left \langle T_{\tau }\psi _{\alpha }^{t}\left(
\mathbf{r}\tau \right) \psi _{\beta }^{s}\left( \mathbf{r}^{\prime }\tau
^{\prime }\right) \right \rangle =-\varepsilon ^{ts}f_{\alpha \beta }\left(
\mathbf{r-r}^{\prime },\tau -\tau ^{\prime }\right) ;  \notag \\
F_{\alpha \beta }^{+ts}\left( \mathbf{r}\tau ,\mathbf{r}^{\prime }\tau
^{\prime }\right) &=&\left \langle T_{\tau }\psi _{\alpha }^{t+}\left(
\mathbf{r}\tau \right) \psi _{\beta }^{s+}\left( \mathbf{r}^{\prime }\tau
^{\prime }\right) \right \rangle =\varepsilon ^{ts}f_{\alpha \beta
}^{+}\left( \mathbf{r-r}^{\prime },\tau -\tau ^{\prime }\right) .  \notag
\end{eqnarray}%
where $t,s$ are the spin indexes. The set of Gor'kov equations in the time
translation invariant, yet inhomogeneous case is\cite%
{Rosenstein17,Rosenstein18},

\begin{eqnarray}
L_{\gamma \beta }^{1}g_{\beta \kappa }\left( \mathbf{r,r}^{\prime }\ \omega
\right) &=&\delta ^{\gamma \kappa }\delta \left( \mathbf{r-r}^{\prime
}\right) -\Delta _{\alpha \gamma }\left( \mathbf{r,}\tau =0\right) f_{\alpha
\kappa }^{+}\left( \mathbf{r,r}^{\prime },\omega \right) ;  \label{FGE} \\
L_{\gamma \beta }^{2}f_{\beta \kappa }^{+}\left( \mathbf{r,r}^{\prime
},\omega \right) &=&\Delta _{\beta \gamma }^{\ast }\left( \mathbf{r,}\tau
=0\right) g_{\beta \kappa }\left( \mathbf{r,r}^{\prime },\omega \right)
\text{.}  \notag
\end{eqnarray}

Here the two Weyl operators are, (tilt vector $\mathbf{w}$ is assumed to be
directed along $x$- axes)

\bigskip
\begin{eqnarray}
L_{\gamma \beta }^{1} &=&\left[ \left( i\omega +\mu ^{\prime }+iw\nabla
_{x}\right) \delta _{\gamma \beta }-iv\sigma _{\gamma \beta }^{i}\ \nabla
_{r}^{i}\right] ;  \label{Eq.12} \\
L_{\gamma \beta }^{2} &=&\left[ \left( -i\omega +\mu ^{\prime }+iw\nabla
_{x}\right) \delta _{\gamma \beta }-iv\sigma _{\gamma \beta }^{it}\nabla
_{r}^{i}\right] \text{,}  \notag
\end{eqnarray}%
Here $\mu ^{\prime }=\mu -\frac{p_{z}^{2}}{2m_{z}}.$

The gap function defined as\bigskip
\begin{equation}
\Delta _{\beta \kappa }^{\ast }\left( \mathbf{r}\right) =g^{2}T\sum
\limits_{\omega }f_{\beta \kappa }^{+}\left( \mathbf{r,}\omega \right) .
\label{Eq. 20}
\end{equation}%
The gap function in the s-wave channel is $\Delta _{\alpha \gamma }\left(
\mathbf{r}\right) =\sigma _{\alpha \gamma }^{x}\Delta \left( \mathbf{r}%
\right) .$ This is the starting point for derivation of the GL free energy
functional of $\Delta \left( \mathbf{r}\right) $.

\section{Derivation of GL equations (without magnetic field)}

In this section the Ginzburg - Landau equations in a homogeneous material
(including the gradient terms) is derived. Magnetic field and fluctuations
effects will be discussed in the next two section by generalizing the basic
formalism.

\subsection{The integral form the Gorkov equations}

To derive the GL equations including the derivative term one needs the
integral form of the Gor'kov equations (see Appendix A), Eq.(\ref{FGE}):
\begin{eqnarray}
g_{\epsilon \kappa }\left( \mathbf{r,r}^{\prime },\omega \right) &=&\mathrm{g%
}_{\epsilon \kappa }^{1}\left( \mathbf{r}-\mathbf{r}^{\prime },\omega
\right) -\int_{\mathbf{r}^{\prime \prime }}\mathrm{g}_{\epsilon \theta
}^{1}\left( \mathbf{r}-\mathbf{r}^{\prime \prime },\omega \right) \Delta
_{\theta \phi }^{\ast }\left( \mathbf{r}^{\prime \prime }\right) f_{\phi
\kappa }^{+}\left( \mathbf{r}^{\prime \prime }\mathbf{,r}^{\prime },\omega
\right) ;  \label{IE1} \\
f_{\beta \kappa }^{+}\left( \mathbf{r,r}^{\prime },\omega \right) &=&\int_{%
\mathbf{r}^{\prime \prime \prime }}\mathrm{g}_{\beta \alpha }^{2}\left(
\mathbf{r-r}^{\prime \prime \prime },-\omega \right) \Delta _{\alpha
\epsilon }^{\ast }\left( \mathbf{r}^{\prime \prime \prime }\right) \times
\notag \\
&&\left \{ \mathrm{g}_{\epsilon \kappa }^{1}\left( \mathbf{r}^{\prime \prime
\prime }-\mathbf{r}^{\prime },\omega \right) -\int_{\mathbf{r}^{\prime
\prime }}\mathrm{g}_{\epsilon \theta }^{1}\left( \mathbf{r}^{\prime \prime }-%
\mathbf{r}^{\prime \prime \prime },\omega \right) \Delta _{\theta \phi
}^{\ast }\left( \mathbf{r}^{\prime \prime }\right) f_{\phi \kappa
}^{+}\left( \mathbf{r}^{\prime \prime }\mathbf{,r}^{\prime },\omega \right)
\right \} \text{.}  \notag
\end{eqnarray}%
Here $\mathrm{g}_{\beta \kappa }^{1}\left( \mathbf{r,r}^{\prime }\right) $
and $\mathrm{g}_{\beta \kappa }^{2}\left( \mathbf{r,r}^{\prime }\right) $
are GF of operators $L_{\gamma \beta }^{1}$ and $L_{\gamma \beta }^{2}$ :

\begin{equation}
L_{\gamma \beta }^{1}\mathrm{g}_{\beta \kappa }^{1}\left( \mathbf{r,r}%
^{\prime }\right) =\delta ^{\gamma \kappa }\delta \left( \mathbf{r-r}%
^{\prime }\right) ;L_{\gamma \beta }^{2}\mathrm{g}_{\beta \kappa }^{2}\left(
\mathbf{r,r}^{\prime }\right) =\delta ^{\gamma \kappa }\delta \left( \mathbf{%
r-r}^{\prime }\right) .  \label{Eq.14}
\end{equation}%
This will be enough do derive the GL expansion to the third order in the gap
function $\Delta \left( \mathbf{r}\right) $ that will be used as an order
parameter\cite{Abrikosov}.

\subsection{The GL expansion}

Using the first and the second iteration of equations Eq.(\ref{IE1}) and
specializing on the case $\mathbf{r}=\mathbf{r}^{\prime }$, one rewrites the
Gorkov's equation Eq.(\ref{FGE}) as (see details in Appendix A):

\begin{equation}
\Delta \left( \mathbf{r}\right) =\frac{g^{2}T}{2}\sum \limits_{\omega }\left
\{ K\left( \mathbf{r-r}_{1}\right) \Delta \left( \mathbf{r}_{1}\right)
-Q\left( \mathbf{r,r}_{1},\mathbf{r}_{2},\mathbf{r}_{3}\right) \Delta \left(
\mathbf{r}_{2}\right) \Delta \left( \mathbf{r}_{3}\right) \Delta \left(
\mathbf{r}_{1}\right) \right \} \text{.}  \label{delta}
\end{equation}%
Here integrations over variables $\mathbf{r}_{1}$, $\mathbf{r}_{2}$, $%
\mathbf{r}_{3}$ are implied. Kernel of the linear in $\Delta $ term is
\begin{equation}
K\left( \mathbf{r}\right) =\mathrm{g}_{21}^{2}\left( \mathbf{r}\right)
\mathrm{g}_{21}^{1}\left( -\mathbf{r}\right) +\mathrm{g}_{11}^{2}\left(
\mathbf{r}\right) \mathrm{g}_{22}^{1}\left( -\mathbf{r}\right) +\mathrm{g}%
_{12}^{2}\left( \mathbf{r}\right) \mathrm{g}_{12}^{1}\left( -\mathbf{r}%
\right) +\mathrm{g}_{22}^{2}\left( \mathbf{r}\right) \mathrm{g}%
_{11}^{1}\left( -\mathbf{r}\right) \text{,}  \label{K}
\end{equation}%
while the coefficient of the cubic term is,

\begin{equation}
Q=%
\begin{array}{c}
\mathrm{g}_{21}^{2}\left( \mathbf{r-r}_{3}\right) \mathrm{g}_{21}^{1}\left(
\mathbf{r}_{2}\mathbf{-r}_{3}\right) \mathrm{g}_{21}^{2}\left( \mathbf{r}_{2}%
\mathbf{-r}_{1}\right) \mathrm{g}_{21}^{1}\left( \mathbf{r}_{1}\mathbf{-r}%
\right) + \\
\ \mathrm{g}_{21}^{2}\left( \mathbf{r-r}_{3}\right) \mathrm{g}%
_{22}^{1}\left( \mathbf{r}_{2}\mathbf{-r}_{3}\right) \mathrm{g}%
_{11}^{2}\left( \mathbf{r}_{2}\mathbf{-r}_{1}\right) \mathrm{g}%
_{21}^{1}\left( \mathbf{r}_{1}\mathbf{-r}\right) + \\
\ \mathrm{g}_{22}^{2}\left( \mathbf{r-r}_{3}\right) \mathrm{g}%
_{11}^{1}\left( \mathbf{r}_{2}\mathbf{-r}_{3}\right) \mathrm{g}%
_{22}^{2}\left( \mathbf{r}_{2}\mathbf{-r}_{1}\right) \mathrm{g}%
_{11}^{1}\left( \mathbf{r}_{1}\mathbf{-r}\right) + \\
\mathrm{g}_{22}^{2}\left( \mathbf{r-r}_{3}\right) \mathrm{g}_{12}^{1}\left(
\mathbf{r}_{2}\mathbf{-r}_{3}\right) \mathrm{g}_{12}^{2}\left( \mathbf{r}_{2}%
\mathbf{-r}_{1}\right) \mathrm{g}_{11}^{1}\left( \mathbf{r}_{1}\mathbf{-r}%
\right) + \\
\mathrm{g}_{11}^{2}\left( \mathbf{r-r}_{3}\right) \mathrm{g}_{21}^{1}\left(
\mathbf{r}_{2}\mathbf{-r}_{3}\right) \mathrm{g}_{21}^{2}\left( \mathbf{r}_{2}%
\mathbf{-r}_{1}\right) \mathrm{g}_{22}^{1}\left( \mathbf{r}_{1}\mathbf{-r}%
\right) + \\
\mathrm{g}_{11}^{2}\left( \mathbf{r-r}_{3}\right) \mathrm{g}_{22}^{1}\left(
\mathbf{r}_{2}\mathbf{-r}_{3}\right) \mathrm{g}_{11}^{2}\left( \mathbf{r}_{2}%
\mathbf{-r}_{1}\right) \mathrm{g}_{22}^{1}\left( \mathbf{r}_{1}\mathbf{-r}%
\right) + \\
\mathrm{g}_{12}^{2}\left( \mathbf{r-r}_{3}\right) \mathrm{g}_{11}^{1}\left(
\mathbf{r}_{2}\mathbf{-r}_{3}\right) \mathrm{g}_{22}^{2}\left( \mathbf{r}_{2}%
\mathbf{-r}_{1}\right) \mathrm{g}_{12}^{1}\left( \mathbf{r}_{1}\mathbf{-r}%
\right) + \\
\mathrm{g}_{12}^{2}\left( \mathbf{r-r}_{3}\right) \mathrm{g}_{12}^{1}\left(
\mathbf{r}_{2}\mathbf{-r}_{3}\right) \mathrm{g}_{12}^{2}\left( \mathbf{r}_{2}%
\mathbf{-r}_{1}\right) \mathrm{g}_{12}^{1}\left( \mathbf{r}_{1}\mathbf{-r}%
\right) \text{.}%
\end{array}
\label{Q}
\end{equation}

Using the Fourier transformation for the GF,

\begin{equation}
\mathrm{g}_{\alpha \beta }^{2,1}\left( \mathbf{r}\right) =\sum \nolimits_{%
\mathbf{p}}g_{\alpha \beta }^{2,1}\left( \mathbf{p}\right) e^{i\mathbf{%
p\cdot r}},\Delta \left( \mathbf{r}\right) =\sum \nolimits_{\mathbf{q}%
}\Delta \left( \mathbf{q}\right) e^{i\mathbf{q\cdot r}}  \label{FT}
\end{equation}%
,

and substituting them into Eqs. (\ref{K}) and (\ref{Q}), one obtains, after
expansion in momenta, the first GL equation,
\begin{equation}
\Delta \left( \mathbf{r}\right) =\frac{g^{2}T}{2}\sum \limits_{\omega ,%
\mathbf{p}}\ \left \{ a\left( \mathbf{p}\right) \Delta \left( \mathbf{r}%
\right) \mathbf{+}C_{ki}\left( \mathbf{p}\right) \frac{\partial ^{2}\Delta
\left( \mathbf{r}\right) }{\partial \mathbf{r}_{i}\partial \mathbf{r}_{k}}%
-b\left( \mathbf{p}\right) \Delta ^{3}\left( \mathbf{r}\right) \right \}
\text{.}  \label{GL17}
\end{equation}%
The function appearing in an expression for the coefficient $a$ is:

\begin{equation}
a\left( \mathbf{p}\right) =g_{21}^{2}\left( \mathbf{p}\right)
g_{21}^{1}\left( \mathbf{p}\right) +g_{11}^{2}\left( \mathbf{p}\right)
g_{22}^{1}\left( \mathbf{p}\right) +g_{12}^{2}\left( \mathbf{p}\right)
g_{12}^{1}\left( \mathbf{p}\right) +g_{22}^{2}\left( \mathbf{p}\right)
g_{11}^{1}\left( \mathbf{p}\right) \text{,}  \label{Eq. 38}
\end{equation}%
while the gradient term coefficients take a form:%
\begin{equation}
C_{ki}\left( \mathbf{p}\right) =\frac{1}{2}\left \{
\begin{array}{c}
\frac{\partial g_{21}^{2}\left( \mathbf{p}\right) }{\partial p_{k}}\frac{%
\partial g_{21}^{1}\left( \mathbf{p}\right) }{\partial p_{i}}+\frac{\partial
g_{11}^{2}\left( \mathbf{p}\right) }{\partial p_{k}}\frac{\partial
g_{22}^{1}\left( \mathbf{p}\right) }{\partial p_{i}}+ \\
\frac{\partial g_{12}^{2}\left( \mathbf{p}\right) }{\partial p_{k}}\frac{%
\partial g_{12}^{1}\left( \mathbf{p}\right) }{\partial p_{i}}+\frac{\partial
g_{22}^{2}\left( \mathbf{p}\right) }{\partial p_{k}}\frac{\partial
g_{11}^{1}\left( \mathbf{p}\right) }{\partial p_{i}}%
\end{array}%
\right \} \text{.}  \label{Eq. 39}
\end{equation}%
The cubic term's coefficient is given by%
\begin{equation}
b\left( \mathbf{p}\right) =\left \{
\begin{array}{c}
g_{21}^{2}\left( \mathbf{p}\right) g_{22}^{1}\left( -\mathbf{p}\right)
g_{11}^{2}\left( -\mathbf{p}\right) g_{21}^{1}\left( \mathbf{p}\right)
+g_{21}^{2}\left( \mathbf{p}\right) g_{21}^{1}\left( -\mathbf{p}\right)
g_{21}^{2}\left( -\mathbf{p}\right) g_{21}^{1}\left( \mathbf{p}\right) + \\
g_{22}^{2}\left( \mathbf{p}\right) g_{11}^{1}\left( -\mathbf{p}\right)
g_{22}^{2}\left( -\mathbf{p}\right) g_{11}^{1}\left( \mathbf{p}\right)
+g_{22}^{2}\left( \mathbf{p}\right) g_{12}^{1}\left( -\mathbf{p}\right)
g_{12}^{2}\left( -\mathbf{p}\right) g_{11}^{1}\left( \mathbf{p}\right) + \\
g_{11}^{2}\left( \mathbf{p}\right) g_{21}^{1}\left( -\mathbf{p}\right)
g_{21}^{2}\left( -\mathbf{p}\right) g_{22}^{1}\left( \mathbf{p}\right)
+g_{11}^{2}\left( \mathbf{p}\right) g_{22}^{1}\left( -\mathbf{p}\right)
g_{11}^{2}\left( -\mathbf{p}\right) g_{22}^{1}\left( \mathbf{p}\right) \ +
\\
g_{12}^{2}\left( \mathbf{p}\right) g_{11}^{1}\left( -\mathbf{p}\right)
g_{22}^{2}\left( -\mathbf{p}\right) g_{12}^{1}\left( \mathbf{p}\right)
+g_{12}^{2}\left( \mathbf{p}\right) g_{12}^{1}\left( -\mathbf{p}\right)
g_{12}^{2}\left( -\mathbf{p}\right) g_{12}^{1}\left( \mathbf{p}\right)%
\end{array}%
\right \} \text{.}  \label{Eq.40}
\end{equation}%
\ The integrations are carried out in the following subsection.

\subsection{Calculation of the coefficients of the GL expansion in a WSM
layer.}

\subsubsection{Linear homogeneous term}

There are two linear in $\Delta $ terms in Eq.(\ref{GL17}). In momentum
space the sum is:%
\begin{equation}
a\left( T\right) =\frac{T}{2}\sum \limits_{\omega ,\mathbf{p}}\ a\left(
\mathbf{p}\right) -\frac{1}{g^{2}}.  \label{a(T)}
\end{equation}%
Substituting the normal GF, calculated in Appendix B for 2D (meaning $p_{z}$
terms in propagators are ignored) is, into Eqs.(\ref{Eq.14},\ref{Eq.12}%
)\bigskip , one obtains coefficient of the linear term,%
\begin{equation}
a\left( \mathbf{p}\right) =2Z^{-1/2}\left \{ \left( vp\right) ^{2}+\omega
^{2}+\left( \mu -w_{x}p_{x}\right) ^{2}\right \} \text{,}  \label{alfa}
\end{equation}%
where

\begin{equation}
\sqrt{Z}=\left( \omega ^{2}+\left( \mu -w_{x}p_{x}-vp\right) ^{2}\right)
\left( \omega ^{2}+\left( \mu -w_{x}p_{x}+vp\right) ^{2}\right) \text{.}
\label{Z}
\end{equation}%
Here and later in the section $\mathbf{p=}\left \{ p_{x},p_{y}\right \} $.

Performing summation on Matsubara frequencies and integration over the 2D
momentum (within the adiabatic approximation, $\mu >>\Omega $, see details
in Appendix C and in \cite{Rosenstein17}) in Eq.(\ref{a(T)}), one obtains:
\begin{equation}
a\left( T\right) =f\ln \frac{T_{c}}{T}\approx \ f\left( 1-\frac{T}{T_{c}}%
\right) .  \label{linear}
\end{equation}%
The critical temperature has the expression (see details in \cite%
{Rosenstein17}) (see Fig.1)
\begin{equation}
T_{c}=1.14\Omega \exp \left[ -1/\lambda \right] ,  \label{Tc}
\end{equation}%
with the effective electron-electron strength in the WSM given by%
\begin{equation*}
\lambda =\lambda _{0}f,\lambda _{0}=\mu g^{2}/2\pi v^{2}\hbar ^{2}\text{.}
\end{equation*}%
The quantity $f$ as a function of the cone tilt parameter $\kappa =w/v$ is
different on the two sides of the topological phase transition of the WSM%
\cite{Rosenstein17}. For the type I WSM, $\kappa <1$, in which the Fermi
surface is a closed ellipsoid, it is given by:%
\begin{equation}
f=\frac{1}{\left( 1-\kappa ^{2}\right) ^{3/2}}\text{.}  \label{f1}
\end{equation}%
\begin{figure}[tbp]
\begin{center}
\includegraphics[width=10cm]{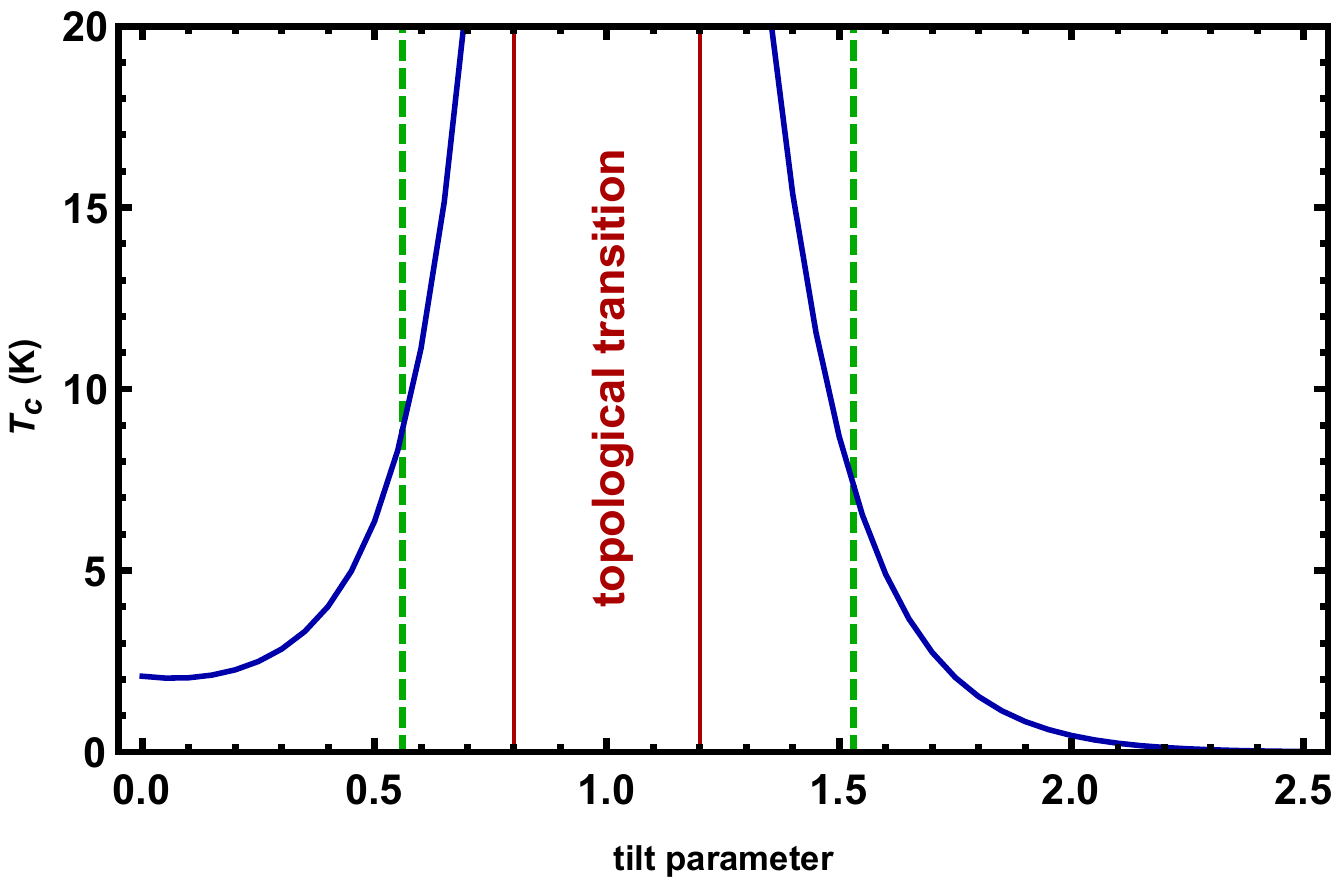}
\end{center}
\par
\vspace{-0.5cm}
\caption{Critical temperature as a function of the tilt parameter $\protect%
\kappa $ indicates Type-I and Type II phases of WSM (Green dashed lines
marks $T_{c}$ for two topological phases of $MoTe_{2}$). Red lines mark the
range where the BCS approximation is not valid. }
\end{figure}
In the type II phase, $\kappa >1$, the Fermi surface becomes open, extending
over the Brillouin zone, and the corresponding expression is:%
\begin{equation}
f=\frac{\kappa ^{2}}{\ \pi \left( \kappa ^{2}-1\right) ^{3/2}}\left \{ 2%
\sqrt{1+\kappa }-1+\log \left[ \frac{2\left( \kappa ^{2}-1\right) }{\kappa
\left( 1+\sqrt{1+\kappa }\right) ^{2}\delta }\right] \right \} \text{.}
\label{f2}
\end{equation}%
Here $\delta $ is an ultraviolet cut off parameter $\delta =a\Omega /w\pi $,
where $a$ is an interatomic spacing. These expression appear in all the
physical quantities calculated below expressing the topological phase
transition. Let us now turn to the gradient terms.

\subsubsection{The gradient terms}

Components $C_{xy}$ and $C_{yx}$ of the second derivative tensor $C$ are
zero due to the reflection symmetry in $p_{y}$ direction , when the cone
tilt vector $\mathbf{w}$ is directed along the $x$ axis (see Appendix D for
details). After integration over momenta in the second term in equation Eq.(%
\ref{GL17}), the gradient terms coefficients are,

\begin{equation}
C_{xx}=\frac{v^{2}\hslash ^{2}}{T_{c}^{2}}\eta _{x},\text{ \ }C_{yy}=\frac{%
v^{2}\hslash ^{2}}{T_{c}^{2}}\eta _{y}\text{,}  \label{CxxCyy}
\end{equation}%
where dimensionless integrals $\eta _{x}$ and $\eta _{y}$ are given in Eqs.(%
\ref{D5},\ref{D6}) of Appendix D.

\subsubsection{Cubic term}

The coefficient of a term cubic in $\Delta $ in the GL equation Eq.(\ref%
{GL17}) reads:

\begin{equation}
b\left( \mathbf{p}\right) =2Z^{-1}\left \{ \left( vp\right) ^{2}\ +\omega
^{2}+\left( \mu -w_{x}p_{x}\right) ^{2}\right \} \left \{ \left( vp\right)
^{2}\ +\omega ^{2}+\left( \mu +w_{x}p_{x}\right) ^{2}\right \} \text{.}
\label{Eq. 44}
\end{equation}%
After integration over momentum, the GL coefficient is obtained%
\begin{equation}
\beta =\frac{\eta }{\mu T_{c}},  \label{betta}
\end{equation}%
with $\eta $ given in Appendix D, Eq.(\ref{D8}). Having determined the
coefficients of the GL equations, we now turn to discussion of the coherence
lengths and the resulting in - plane anisotropy due to the tilt of the Dirac
cone.

\subsection{In plane coherence lengths and anisotropy}

\subsubsection{Coherence lengths}

The first GL equation in WSM in magnetic field (required in the following
section) is standard:

\begin{equation}
-\left( \xi _{x}^{2}\partial _{x}^{2}+\xi _{y}^{2}\partial _{y}^{2}\right)
\Delta \left( \mathbf{r}\right) -\tau \Delta \left( \mathbf{r}\right) +\frac{%
\beta }{f}\left \vert \Delta \left( \mathbf{r}\right) \right \vert
^{2}\Delta \left( \mathbf{r}\right) =0\text{.}  \label{GLH}
\end{equation}%
Here $\tau =1-T/T_{c}$. Comparing coefficients of linear terms in Eq.(\ref%
{GLH}), the coherence lengths are
\begin{equation}
\xi _{x}^{2}=C_{xx}/f,\xi _{y}^{2}=C_{yy}/f\text{,}  \label{ksi}
\end{equation}%
and are computed numerically.

To be specific the in plane correlations lengths are calculated for a $%
MoTe_{2}$ single crystals that were extensively studied experimentally at
pressures between ambient to $30GPa$. The coherence lengths $\xi _{x}$ and $%
\xi _{y}$ as functions of the tilt ration $\kappa =w/v$ for material
parameters pertinent to $MoTe_{2}$ are shown in Fig. 1 as solid blue and
green lines respectively. We estimate the Debye frequency from the Raman data%
\cite{MoTe2melting}, $\Omega =100K$. Fermi velocity $v=5\cdot 10^{7}cm/s$
and Fermi energy, $\mu =8\Omega $, from ARPES\cite{MoTeearly}. An
ultraviolet cutoff for Eq.(\ref{f2}) is taken to be an interatomic distance $%
a=0.3nm$ ($T_{c}$ depends logarithmically on it, see Eq.(\ref{f2}). The
electron - electron coupling due to phonons $\lambda _{0}=g^{2}\mu /2\pi
v^{2}\hbar ^{2}$ is assumed to be linearly dependent of $\kappa $ (or
pressure that presumably determines $\kappa $): $\lambda _{0}=\lambda
_{0}^{I}-\alpha \kappa $ for $\lambda _{0}^{I}=0.25$ and $\alpha =0.05$.

One observes that the both coherence lengths are large and roughly equal at
small $\kappa $. Below $\kappa =0.2$ the curve flattens reaching value of $%
\xi _{x}=\xi _{y}=45nm$ for graphene - like material at $\kappa =0$. In the
topological transition region (marked in Fig.2a by red lines) they become
very small. In the type II phase the two coherence length are different and
become large again. \thinspace In the critical region the theory becomes
inapplicable.\

\subsubsection{In plane anisotropy}

The anisotropy parameter is defined as $\varepsilon =\xi _{x}/\xi _{y}=\sqrt{%
C_{xx}/C_{yy}}$. It is plotted as a function of $\kappa $ in Fig.2b. The
coherence length in $z-$direction $\xi _{z}$ as a function of tilt parameter
$\kappa $ is presented in Fig.2c.

\begin{figure}[tbp]
\begin{center}
\includegraphics[width=18cm]{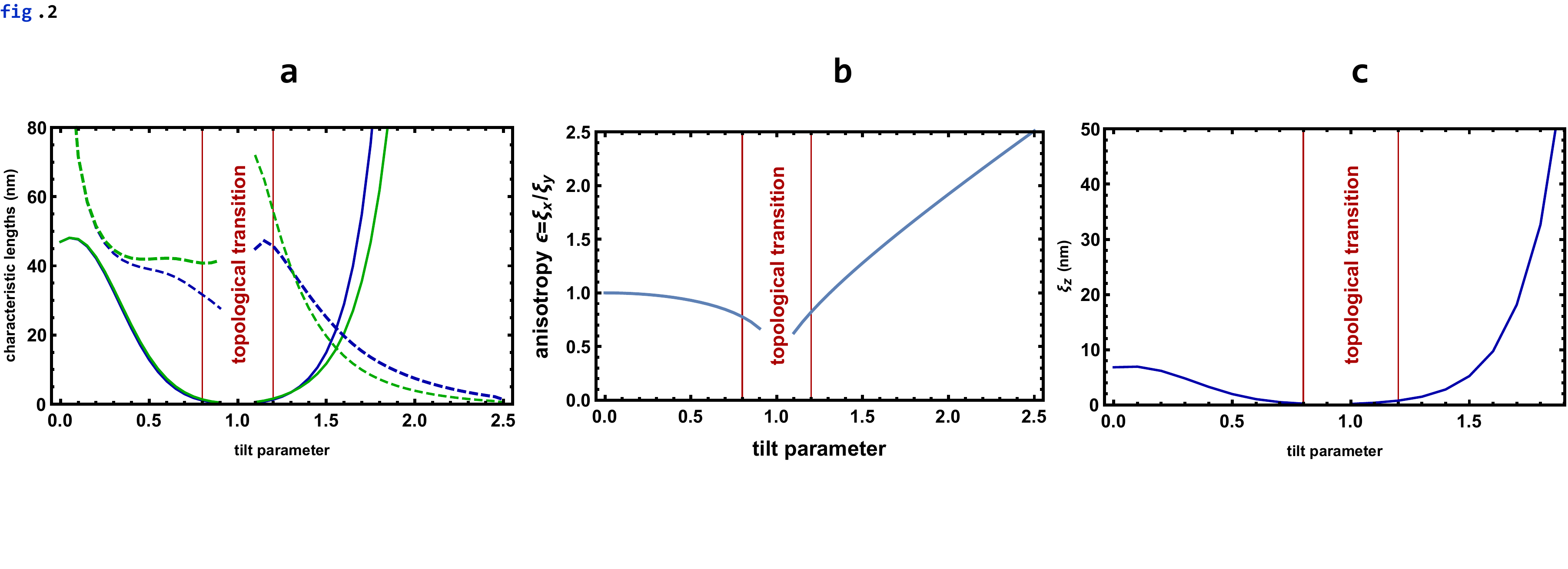}
\end{center}
\par
\vspace{-0.5cm}
\caption{a. Dependence of characteristic lengths of the Weyl superconductor
on the tilt parameter $\protect\kappa $. The topological (Lifshitz)
transition occurs at $\protect\kappa \rightarrow 1$. Coherence lengths along
the $x$ (blue) and $y$ (green) directions are solid lines. Same for the
penetration depth times $\protect\sqrt{2}$ as dashed lines. b. In-plane
anisotropy of the coherence length $\protect\xi _{x}/\protect\xi _{y}$ (same
as the ratio of penetration depths $\protect\lambda _{y}/\protect\lambda %
_{x} $) \ as function of the tilt parameter. c. Characteristic length $%
\protect\xi _{z}$ in direction perpendicular to the layers on the tilt
parameter $\protect\kappa $. Here the thickness of single layer $s=3nm,$ and
interlayer distance $d=10nm.$}
\label{Fig.2}
\end{figure}

Graphene - like superconductor is isotropic. At small $\kappa $ the
anisotropy is small with $\epsilon <1$. Above the topological phase
transition line it increases rapidly with $\kappa >1$ and becomes much
larger than $1$ already at $\kappa =1.2$. Unfortunately there is no known
purely WSM superconducting 2D material at this time and therefore we
consider a 3D material with similar properties.

\section{Layered WSM}

Till now a single 2D layer was considered. The stack of these layers, see
Fig.3, forms the 3D WSM dichalcogenides like $MoTe_{2}$. In these systems
the thin superconducting layers (thickness $s$) are separated by distance $d$
and are bound by the Van der Waals interaction. In order to calculate GL
expansion coefficients in this case we \ use the perturbation on the
effective mass $m_{z}$ procedure when the set of the 2D nonbounded layers
are considered as the zero approximation in perturbation theory. Parabolic
term of the Hamiltonian responsible for interlayers interaction should be
taken into account to calculate the GL expansion coefficient in $z$
direction $C_{zz}\frac{d^{2}\Delta }{dz^{2}}$. In this case one has to
perform 3D Fourier transformation in Eq. \ref{FT} while 2D vectors $\mathbf{r%
}$ should be replaced by 3D vector $\mathbf{r}=\left( x,y,z\right) $. The 3D
momentum in this case is $\left( \mathbf{p},p_{z}\right) $.

The GL expansion in Eq. \ref{GL17} has the same form as in 2D case with
additional gradient term in $z$ direction $C_{zz}\frac{d^{2}\Delta }{dz^{2}}$
while the chemical potential $\mu $ should be replaced by $\mu -\frac{%
p_{z}^{2}}{2m_{z}}$ in all of the GF. The 3D integration over momentum in
this case gives (see details in Appendix D),
\begin{equation}
C_{zz}=\frac{\hbar s}{2\pi ^{2}\mu }\sqrt{\frac{T_{c}}{2m_{z}}}\eta _{z}%
\text{,}  \label{Czz}
\end{equation}%
where $\eta _{z}$ is the dimensionless function depending on the chemical
potential $\mu $ and the tilt parameter $\kappa $. The coherence length for $%
MoTe_{2}$ in the $z$-direction $\xi _{z}^{2}=C_{zz}/f$, is presented in
Fig.2c. (We have found by direct calculation that the function $f$ does not
change when we extend to 3D). Calculations of effects of magnetic field and
thermal fluctuations require the GL free energy.

\subsection{Free GL energy for layered WSM superconductor}

The corresponding Ginzburg-Landau functional now has a form:

\begin{equation}
F=\int d^{3}r\text{ }D\left( \mu \right) \left( \xi _{i}^{2}\left \vert
\partial _{i}\Delta \left( \mathbf{r}\right) \right \vert ^{2}-\tau \left
\vert \Delta \left( \mathbf{r}\right) \right \vert ^{2}+\frac{\beta }{2f}%
\left \vert \Delta \left( \mathbf{r}\right) \right \vert ^{4}\right) \text{.}
\label{F0}
\end{equation}%
where $i=x,y,z.$ Here (see Appendix E) $D\left( \mu \right) $ is the one
particle density of states (DOS) for WSM with arbitrary cone slope parameter
$\kappa $, Eq.(\ref{eq1}),
\begin{equation}
D\left( \mu \right) =D_{0}\left( \mu \right) f\   \label{DOS17}
\end{equation}%
where $D_{0}\left( \mu \right) =\sqrt{2m_{z}}\mu ^{3/2}/3\pi ^{2}\hbar
^{3}v^{2}$ is DOS for layered "graphene" ($\kappa =0$). The GL functional
for layered system consisting on 2D superconducting layered separated by the
dielectric inter-layers incorporates the Josephson coupling. The tunneling
of the electrons moving between the superconducting layers via dielectric
streak described by the effective mass $m_{z}$ of the electrons moving along
the $z$ axis. Within tight binding model the effective mass is estimated as $%
m_{z}=m_{e}s^{2}/d^{2}\exp \left[ d/s\right] $, where $m_{e}$ is the mass of
free electron, $d$ is the distance between layers of thickness $s$, see
Fig.3.

\begin{figure}[tbp]
\begin{center}
\includegraphics[width=16cm]{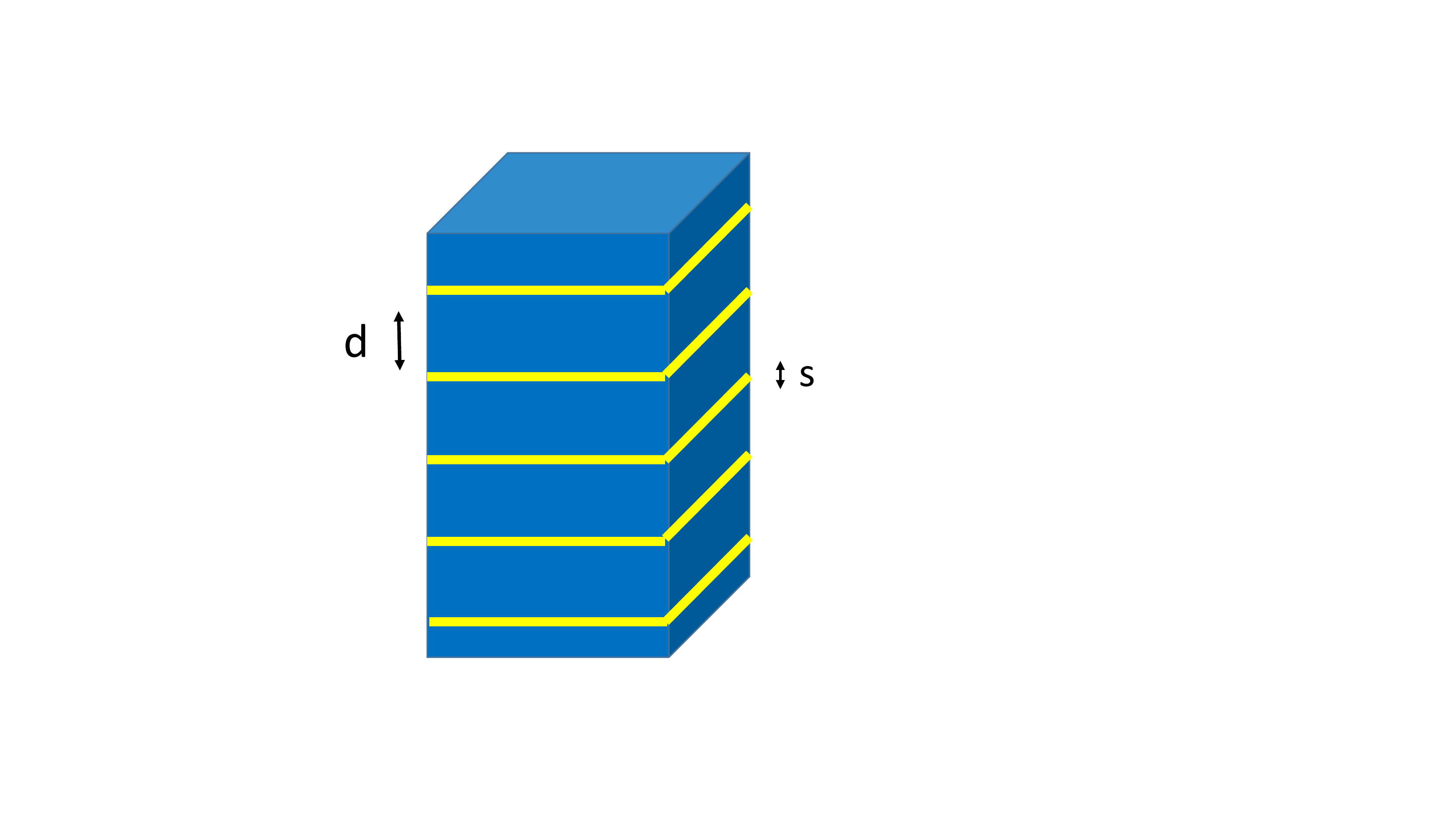}
\end{center}
\par
\vspace{-0.5cm}
\caption{Layered Weyl semi - metal - a schematic picture. }
\end{figure}

Using the equilibrium value of the order parameter,%
\begin{equation}
\Delta ^{2}=\frac{f}{\beta }\tau ,  \label{deltasquare}
\end{equation}%
the condensation energy density of an uniform superconductor (required in
section V to describe thermal fluctuations' importance) is
\begin{equation}
F_{s}=\int d^{3}r\text{ }D\left( \mu \right) \left[ -\tau \left \vert \Delta
\right \vert ^{2}+\frac{\beta }{2f}\left \vert \Delta \right \vert ^{4}%
\right] =-\frac{D_{0}\left( \mu \right) f\Delta ^{2}\tau }{2}V=-\frac{%
D_{0}\left( \mu \right) f^{2}}{2\beta }\tau ^{2}V.  \label{energy}
\end{equation}%
Now we are ready to describe the magnetic properties of the superconductors.$%
\ \ \ $

\section{GL in magnetic field. Comparison with experiment.}

Effects of the external magnetic field are accounted for by the minimal
substitution, $\mathbf{\nabla }\rightarrow \mathbf{D}=\mathbf{\nabla }-\frac{%
2ei}{c}\mathbf{A}$ in the GL equation Eq.(\ref{GLH}) due to gauge invariance.%
$\ $The GL equation in the presence of magnetic field allows the description
of the magnetic response to homogeneous external field. We start from the
strong field that destroys superconductivity.

\subsection{Upper critical field.}

The upper critical magnetic field $H_{c2}$ is as usual calculated from the
liner part of the GL equation Eq.(\ref{GLH}), as the lowest eigenvalue of
the linear operator (including the magnetic field). Representing the
homogeneous magnetic field in the Landau gauge, $A=H\left( -y,0,0\right) $,
\bigskip one expands near $T_{c}$ as%
\begin{equation}
H_{c2}\left( T\right) =H_{c2}\left( 0\right) \tau ,\   \label{Eq. 64}
\end{equation}%
where the zero temperature intercept magnetic field is $H_{c2}\left(
0\right) =\Phi _{0}/2\pi \xi _{x}\xi _{y}$. This is represented by$\ $the
dashed straight lines in Fig. 4. It is a product of the experimentally
measured slope $\frac{dH_{c2}}{dT}|_{_{T=T_{c}}}$ and $T_{c}$:
\begin{equation}
H_{c2}\left( 0\right) =\frac{\hbar cf}{2e\sqrt{C_{yy}C_{xx}}}.
\label{Eq. 65}
\end{equation}%
In practice at very low temperature the mean field $H_{c2}\left( T\right) $
"curved down", so that actual upper field at zero temperature is about $60\%$
of that value. The GL model is not applicable that far from $T_{c}$.

\begin{figure}[tbp]
\begin{center}
\includegraphics[width=10cm]{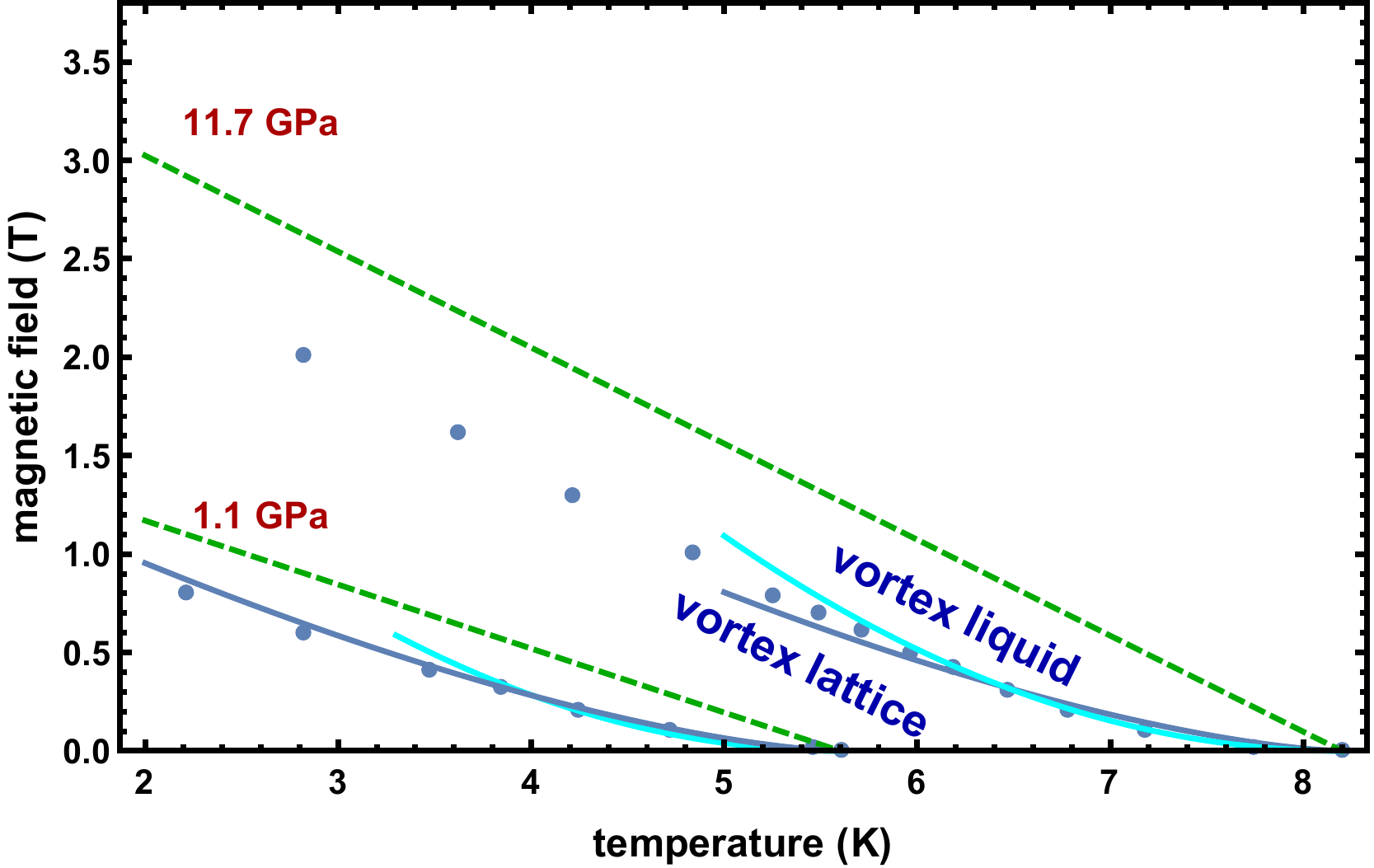}
\end{center}
\par
\vspace{-0.5cm}
\caption{Magnetic phase diagram of layered WSM second kind superconductor.
The experimental points are for $MoTe_{2}$ at pressures \thinspace $1.1GPa$
and $11.7GPa$ (blue). Upper critical field $H_{c2}\left( T\right) $ ("mean
field",dashed line) becomes a crossover due to thermal fluctuations.$\ $At
pressures $1.1$ and $11.7GPa$ the fitted curves are marked by the cyan lines
for 3D and the blues line for 2D.}
\end{figure}

Measured upper critical field as function for parameter of $MoTe_{2}$ for
two values of pressure, $1.1$ $GPa$ and $11.7$ $GPa$, is given as a red and
blue points respectively line in Fig. 4. As will be discussed below, it will
be interpreted as a melting line for the vortex lattice due to fluctuations.
Vortex liquid phase in which the phase of the order parameter $\Delta $ is
random appears between the melting line and the mean field line where order
parameter disappears altogether.

Pressure determines the tilt parameter $\kappa $, which in turn influences $%
H_{c2}\left( 0\right) $, as shown in Fig. 5 (blue lines). In the
superconductor of the first kind it becomes the cooling field and is
depicted as dashed lines at both small and large $\kappa $.

\begin{figure}[tbp]
\begin{center}
\includegraphics[width=10cm]{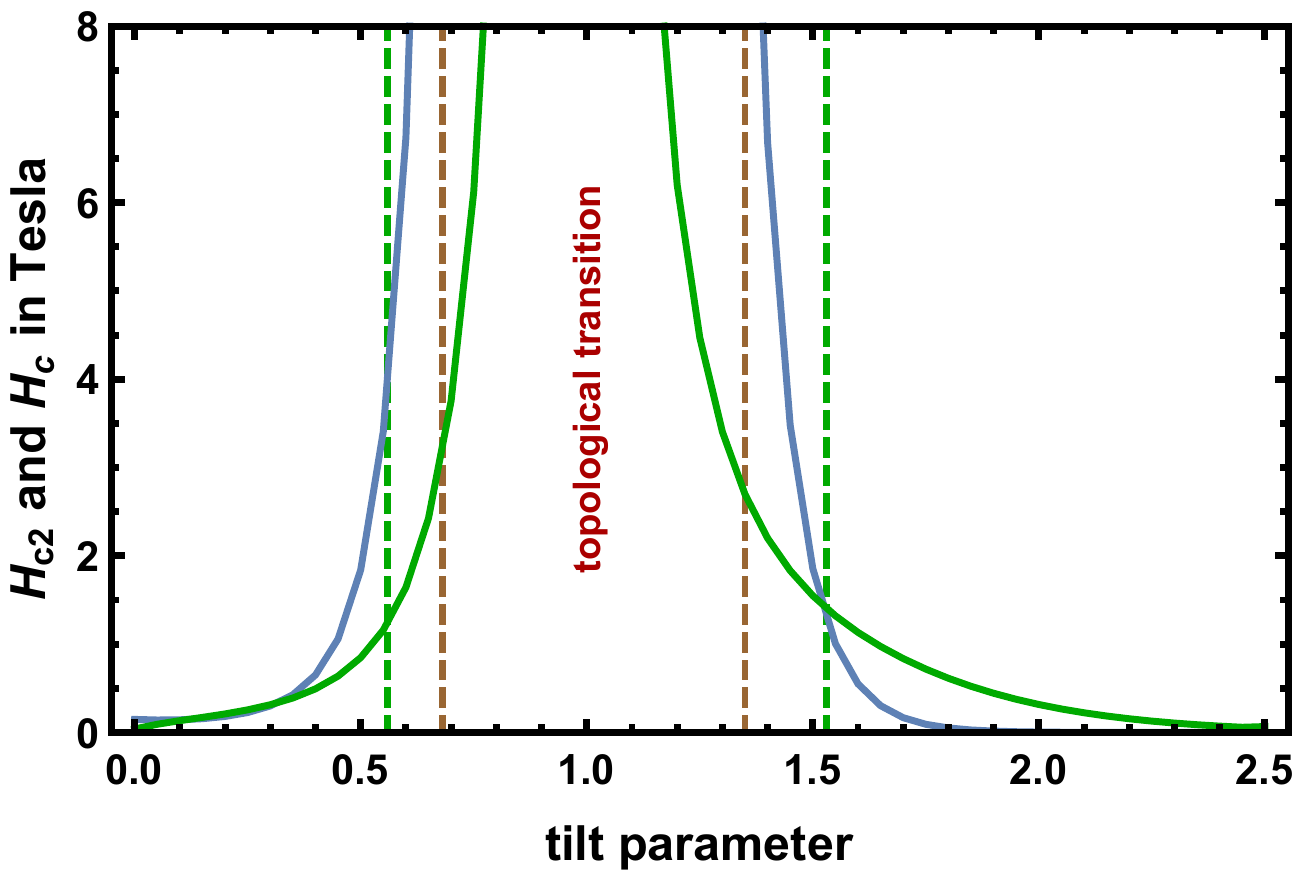}
\end{center}
\par
\vspace{-0.5cm}
\caption{Upper critical field ($H_{c2}$, blue lines for type I and type II
phases) and thermodynamic critical magnetic field ($H_{c}$, green) as a
function on the tilt parameter. Brown dashed lines mark $T_{c}$ for two
topological phases of $MoTe_{2}$.}
\end{figure}

\subsection{Supercurrents and penetration depths in London limit.}

\subsubsection{Penetration depth.}

Density of superconducting currents can be obtained by the variation of the\
free energy functional including the magnetic energy,
\begin{equation}
F=\int d^{3}r\left \{ D_{0}\left( \mu \right) \left( C_{ii}\left \vert
D_{i}\Delta \right \vert ^{2}-f\tau \left \vert \Delta \right \vert ^{2}+%
\frac{\beta }{2}\left \vert \Delta \right \vert ^{4}\right) +\frac{\left(
\nabla \times \mathbf{A}\right) ^{2}}{8\pi }\right \} \text{,}
\label{energy3D}
\end{equation}%
where $i=x,y,z$, with respect to components of the vector potential:\
\begin{equation}
\mathbf{J}_{i}=D_{0}\left( \mu \right) \frac{2ei}{\hbar }C_{ii}\Delta \left(
\mathbf{r}\right) D_{i}\Delta ^{\ast }\left( \mathbf{r}\right) +c.c.\text{.}
\label{j}
\end{equation}%
Within the London approximation, in which the order parameter is
approximated by $\Delta \left( \mathbf{r}\right) =\Delta e^{i\varphi }$, one
obtains,

\begin{equation}
\mathbf{J}_{i}=\frac{4e}{\hbar }D_{0}\left( \mu \right) C_{ii}\Delta
^{2}\left( \partial _{i}\varphi -\frac{2e}{c\hbar }A_{i}\right) \text{.}
\label{j1}
\end{equation}%
Using the (in plane) Maxwell equations, one obtains the equation for a
single Abrikosov vortex \cite{Keterson}:
\begin{equation}
\lambda _{x}^{2}\left( T\right) \frac{\partial ^{2}H}{\partial y^{2}}%
+\lambda _{y}^{2}\left( T\right) \frac{\partial ^{2}H}{\partial x^{2}}-H=%
\frac{\Phi _{0}}{2\pi }\delta \left( x\right) \delta \left( y\right) \text{.}
\label{Lamda}
\end{equation}%
The London penetration lengths in our case of layered WSM with parabolic
dispersion relation along $z$ axis are:
\begin{equation}
\lambda _{x}^{2}\left( T\right) =\frac{c^{2}\hbar ^{2}}{32\pi
e^{2}D_{0}\left( \mu \right) C_{yy}\Delta ^{2}}\text{.}  \label{lambdas}
\end{equation}

From the calculated coefficient of the cubic term of the GL equation and the
Maxwell equation one obtains, after substitution of $D\left( \mu \right) $
from Eq.(\ref{DOS17}) and $\Delta $ from Eq.(\ref{deltasquare}),

\begin{equation}
\lambda _{x}^{2}\left( 0\right) =\frac{3\pi \hbar ^{5}v^{2}c^{2}\beta }{32%
\sqrt{2}e^{2}m_{z}^{1/2}\mu ^{3/2}C_{yy}f}\text{, }\lambda _{y}=\lambda
_{x}/\varepsilon \text{.}  \label{L0x}
\end{equation}%
The quantities $\sqrt{2}\lambda _{x}\left( 0\right) $ and $\sqrt{2}\lambda
_{y}\left( 0\right) $ are depicted in Fig.2a as dashed blue and green lines
respectively. The factor $\sqrt{2}$ was introduced in order to mark the
transitions from the first to second kind of superconductivity. For material
parameters used in the present paper ($MoTe_{2}$) the transitions are
reentrant in $\kappa $: $\kappa _{I}=0.53$ and $\kappa _{II}=1.5$
(intersection points with $\xi _{x}$ or consistently with $\xi _{y}$). The
parameters that determine $m_{z}$ (see formula below Eq.(\ref{DOS17})), are
the interlayer distance $d=1.3nm$, the layer effective width $s=0.3nm$. The
dependence is quite non-monotonic. At small $\kappa $ both penetration
depths are large level off and increase slightly approaching $\kappa =1$. In
the type II phase penetration depth largely decreases.

\subsubsection{The Abrikosov parameter and transition between first and
second kinds of superconductivity}

The Abrikosov parameter is isotropic despite large anisotropies:

\begin{equation}
\kappa _{x}^{A}=\frac{\lambda _{x}}{\xi _{x}}=\frac{vc}{8e}\sqrt{\frac{3%
\sqrt{2}\pi \hbar ^{5}\beta }{m_{z}^{1/2}\mu ^{3/2}C_{xx}C_{yy}}}=\kappa
_{y}^{A}\ \text{.}  \label{Eq.69}
\end{equation}%
This is plotted against the tilt parameter in Fig. 6. The green line is the
universal critical value $\kappa _{x}^{A}=1/\sqrt{2}$ for the above
mentioned transitions between the first and the second kind
superconductivity.

\begin{figure}[tbp]
\begin{center}
\includegraphics[width=10cm]{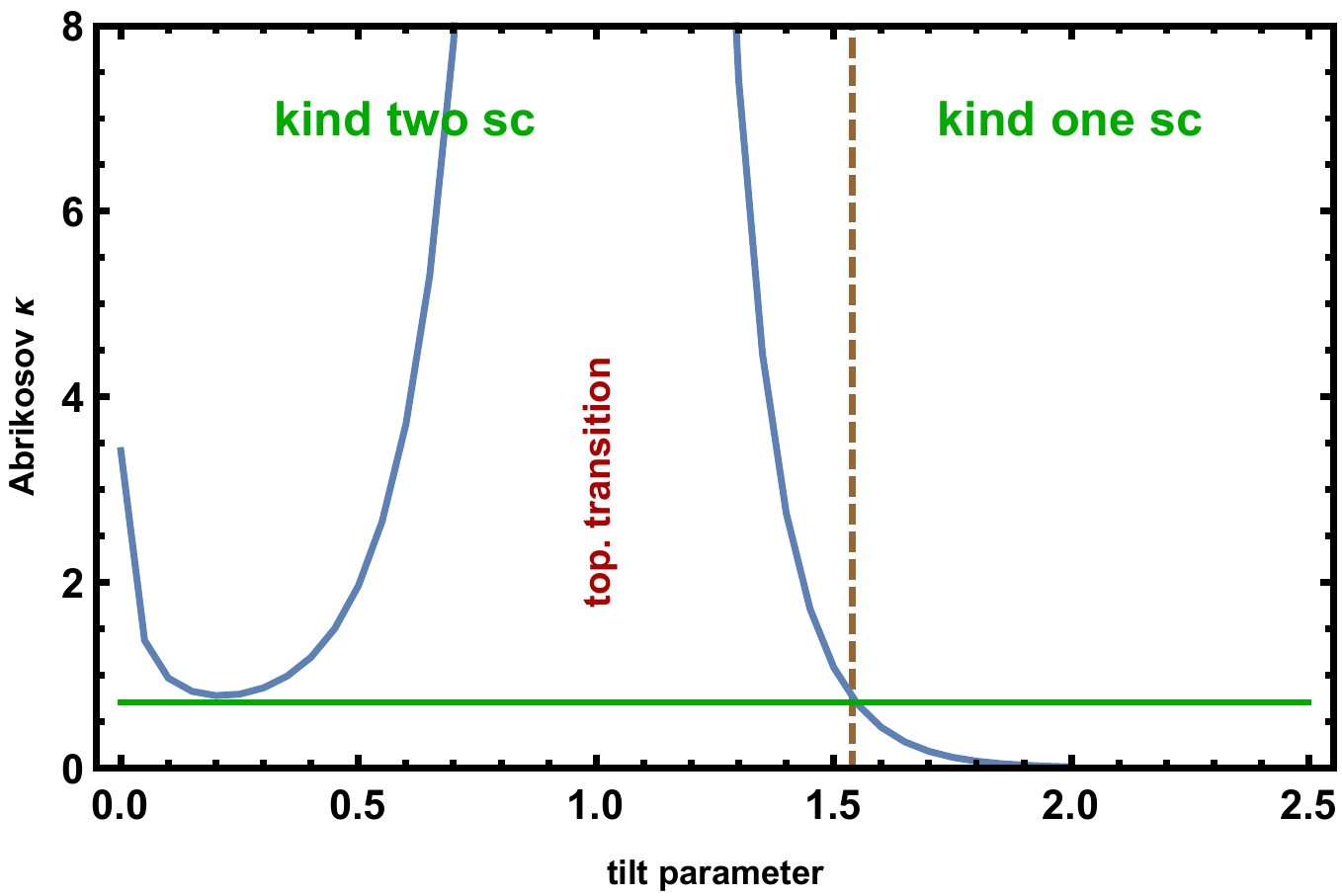}
\end{center}
\par
\vspace{-0.5cm}
\caption{Abrikosov parameter of the WSM superconductor as function of $%
\protect\kappa $. The green line is the universal critical value $\protect%
\kappa _{x}^{A}=1/\protect\sqrt{2}$for the transitions between the first and
the second kind superconductivity.}
\end{figure}
Thermodynamic critical field for kind I superconductors is given by

\begin{equation}
H_{c}^{2}\left( 0\right) =8\pi F_{s}=4\pi D_{0}\left( \mu \right) f\Delta
^{2}=\frac{4\sqrt{2m_{z}}\mu ^{3/2}f^{2}}{3\pi \hbar ^{3}v^{2}\beta }\text{,}
\label{Hc}
\end{equation}%
where the condensation energy was given in Eq.(\ref{energy}). It is plotted
as dashed lines in Fig.5 as dashed lines.

\subsection{The Abrikosov vortex solution and the lower critical field.}

In a hard type-II superconductor magnetic field screened the Abrikosov
vortex obeyed the equation Eq.(\ref{Lamda}). This equation has a well known
anisotropic Abrikosov vortex solution\cite{Keterson}:
\begin{equation}
H\left( x,y\right) =\frac{\Phi _{0}}{2\pi \lambda _{x}\lambda _{y}}K_{0}%
\left[ \left( \frac{y^{2}}{\lambda _{x}^{2}}\ +\frac{x^{2}}{\lambda _{y}^{2}}%
\right) ^{1/2}\right] \text{.}  \label{Eq.72}
\end{equation}%
here $K_{0}$ is the modified Bessel function.

Abrikosov vortex in WSM appears at lower critical field
\begin{equation}
H_{c1}\left( 0\right) =\frac{\Phi _{0}}{2\pi \lambda _{x}\lambda _{y}}\ln %
\left[ \kappa ^{A}\right] \text{.}  \label{Eq.73}
\end{equation}

\bigskip The material parameters calculated above allow determination of the
strength of thermal fluctuations that might be significant in thin films as
seen from the nonlinear concave shape of measured\cite{MoTe2melting}
transition field dependence on temperature near $T_{c}$ in $MoTe_{2}$
superconductor, see Fig. 4. However the experimental points of the magnetic $%
H_{c2}$ (blue dots in Fig.4) indicate that the mean field description breaks
down near $T_{c}$. This will be explained next as a thermal fluctuations
effect.

\section{Ginzburg criterion for strong thermal fluctuations region.}

The thermal fluctuations were neglected so far. In this section they are
taken into account in the framework of the GL energy. Here one cannot ignore
the fluctuations of the order parameter in direction perpendicular to the
layer, since magnetic field couples the layers via the "pancake vortices"
interaction\cite{RMP}.

\subsection{Ginzburg number in layered superconductor}

The fluctuation contribution to the heat capacity (per volume) that is most
singular in $\tau =1-T/T_{c}$ is\cite{Patashinsky,Landau}:
\begin{equation}
C_{fluct}=\frac{\pi ^{2}}{\xi _{x}\xi _{y}\xi _{z}}\frac{1}{\sqrt{\tau }}%
\text{.}  \label{fl2}
\end{equation}%
It should be compared with the mean field heat capacity $C_{mf}$ in the
superconducting phase (see Eqs.(\ref{deltasquare}) and (\ref{DOS17}):
\begin{equation}
C_{mf}=\frac{D_{0}\left( \mu \right) f^{2}}{\beta T_{c}}=\frac{\sqrt{2m_{z}}%
\mu ^{3/2}}{3\pi ^{2}\hbar ^{3}v^{2}}\frac{f^{2}}{\beta T_{c}}\text{.}
\label{Cmf}
\end{equation}%
The ratio,%
\begin{equation}
\frac{C_{fl}}{C_{mf}}=\frac{3\pi ^{4}\hbar ^{3}v^{2}}{\sqrt{2m_{z}\mu ^{3}f}}%
\frac{\beta T_{c}}{\sqrt{C_{xx}C_{yy}C_{zz}}}\frac{1}{\sqrt{\tau }}\text{,}
\label{ratio}
\end{equation}%
characterizes the fluctuation strength. Strong fluctuations effects appear
in the temperature region where $C_{fl}>C_{mf}$. The temperature independent
Levanyuk - Ginzburg number is defined by:

\begin{equation}
Gi^{th}=\frac{9\pi ^{8}\hbar ^{6}v^{4}}{2m_{z}\mu ^{3}}\frac{\beta
^{2}T_{c}^{2}}{C_{xx}C_{yy}C_{zz}f}\text{.}  \label{Gi}
\end{equation}

The Ginzburg number is plotted as function of $\kappa $ in Fig.7. for
parameters pertinent to an experiment\cite{MoTe2melting} in $MoTe_{2}$. In
this case $Gi$ ranges between relatively large values in Type I WSM phase $%
\kappa $ close to the topological transition line and small $Gi$ value in
Type II WSM phase. In type I phase there exists a minimum. Significant
thermal fluctuations lead to melting of the Abrikosov flux lattice to the
vortex liquid. Values of $Gi$ for $MoTe_{2}$ at pressures $1.1GPa$ and $%
11.7GPa$ clearly exhibiting the melting line\cite{MoTe2melting} are given in
Table 1.

\begin{figure}[tbp]
\begin{center}
\includegraphics[width=10cm]{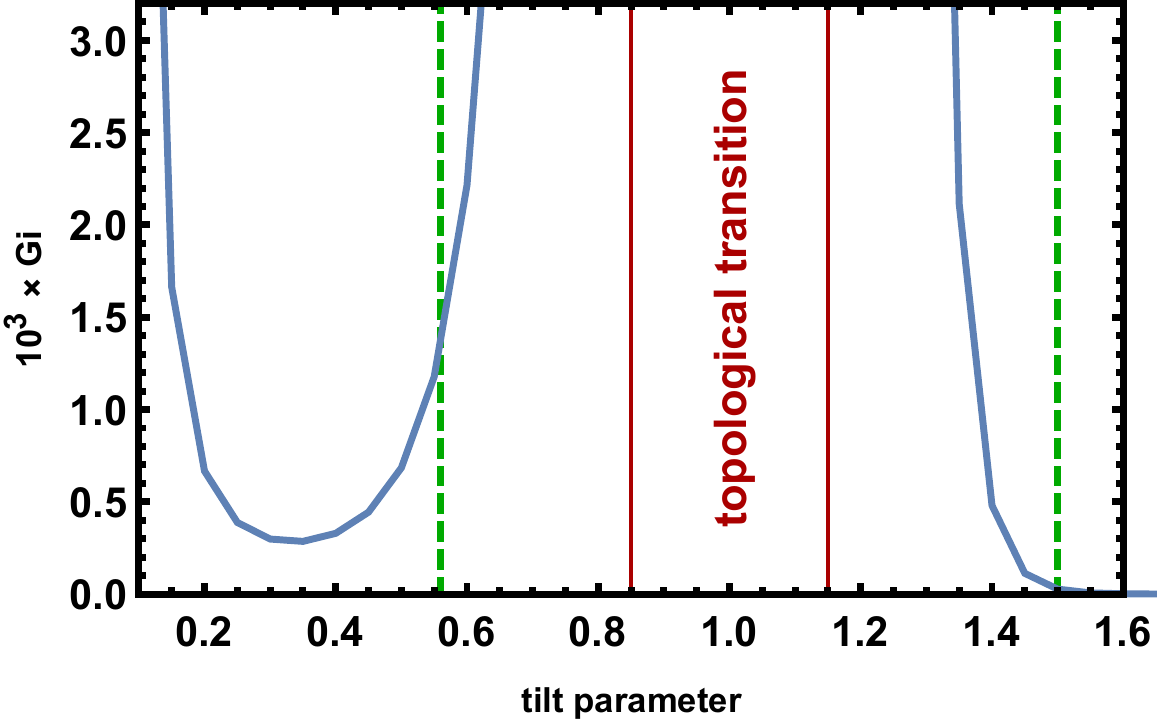}
\end{center}
\par
\vspace{-0.5cm}
\caption{Gi number characterizing the strength of thermal fluctuations as
function of the tilt parameter $\protect\kappa .$}
\end{figure}

\subsection{Abrikosov lattice melting line}

It was shown\cite{Rosenstein02} that the melting line is determined for 3D
and 2D thermal fluctuations\cite{shapiro88} by
\begin{eqnarray}
-a_{T}^{3D} &=&2^{1/3}\left( th\right) ^{-2/3}Gi^{-1/3}\left( 1-t-h\right)
=9.5;  \label{aT} \\
-a_{T}^{2D} &=&2^{-1/4}\left( th\right) ^{-1/2}Gi^{-1/4}\left( 1-t-h\right)
=13.2,  \notag
\end{eqnarray}%
respectively. Here the scaled melting field, see Fig. 4, is $%
h=H/H_{c2}\left( 0\right) \ $\ and $t=T/T_{c}$. The values of Thouless
parameter \cite{RMP} at the first order melting transition were determined
by comparing energies of the vortex solid and liquid found nonperturbatively.

In the vicinity of $T_{c}$, namely for $h,1-t<<1$, the expression for the
melting field simplifies $H_{m}^{D}\left( T\right) =H_{m}^{D}\left(
1-t\right) ^{3-D/2}$ with values of $H_{m}^{D}$ given by
\begin{eqnarray}
H_{m}^{2} &=&\frac{1}{\left( 13.2\right) ^{2}\sqrt{2Gi_{2}}}H_{c2}\left(
0\right)  \label{HD} \\
H_{m}^{3} &=&\frac{\sqrt{2}}{\left( 9.5\right) ^{3/2}\sqrt{Gi_{3}}}%
H_{c2}\left( 0\right)  \notag
\end{eqnarray}

In our case of $MoTe_{2}$ at pressures $1.1$ and $11.7GPa$ the fitted
constants (see the cyan lines for 3D and the blue lines for 2D in Fig.4),
one obtains the best fits for $H_{m}^{D}$ given in Table 1.

\begin{table*}[tph]
\caption{Fitting parameters for $H_{m}^{D}$}
\begin{center}
\renewcommand \arraystretch{1.5}
\begin{tabularx}{\textwidth}{XXXXXXXXXXX}
\hline
\hline
\text{pressure} & $T_{c}$ & $\kappa $  & $H_{c2}\left( 0\right)$  & $\xi _{x}$ &
$\lambda _{x}$ & $H_{m}^{2}$ & $H_{m}^{3}$ & $Gi^{2Dfit}$ & $Gi^{3Dfit}$ & $Gi^{th}$  \\
\hline
1.1GPa & 5.6 & 1.5 & 1.5T & 18nm & 20nm & 3.5T & 1.85T & 3$\cdot 10^{-6}$ &
$1.5\cdot 10^{-3} $& $2.7\cdot 10^{-5}$ \\
11.7GPa & 8.2K & 0.53 & 4T & 10nm & 40nm & 7.2T & 3.3T & $ 5\cdot 10^{-6}$ &
$3.4\cdot 10^{-3} $& $ 1.3\cdot 10^{-3}$ \\
\hline
\hline
\end{tabularx}
\end{center}
\end{table*}

The $Gi$ in both cases was determined from the several experimental points
close to $T_{c}$ using

\begin{eqnarray}
Gi^{D} &=&c_{D}\left( H_{c2}\left( 0\right) /H_{m}\right) ^{2};
\label{Gifit} \\
c_{2} &=&1.64\cdot 10^{-5};c_{3}=2.33\cdot 10^{-3}\text{.}  \notag
\end{eqnarray}%
while $Gi^{th}$ is calculated in Eq.(\ref{Gi}).

The actual melting line significantly below $T_{c}$ typically bends down and
cannot be obtained within the GL expansion. The theoretical value in the
table is taken from Fig.7.

\section{Conclusion and discussion.}

Magnetic properties of Weyl semi - metals turned superconductors at low
temperatures were derived from a microscopic phonon mediated multi - band
pairing model via the Ginzburg - Landau effective theory for the (singlet)
order parameter. The Gorkov approach was used to determine microscopically
anisotropic coherence length, the penetration depth, Fig.2a, determining the
Abrikosov parameter for a layered material. It is shown that very strong in
plane anisotropy is caused by the tilt of Dirac cones, see Fig. 2b. It is
found that generally that superconductivity is strongly second kind
(penetration depth much larger than coherence length) near the WSM
topological transition (tilt parameter $\kappa =1$, see Fig. 6), but becomes
first kind away from it especially in type II WSM. This possibility has been
observed recently in similar material\cite{PdTe2TypeI} $PaTe_{2}$.

For WSM superconductors of the second kind the dependence of the upper and
lower critical fields $H_{c2}\left( T\right) $ and $H_{c1}\left( T\right) $
on the tilt parameter $\kappa $ (governed by pressure, see Fig. 5) was
obtained from the GL energy not very far from $T_{c}$ (where the GL approach
is valid). In WSM superconductors of first kind the relevant fields are the
thermodynamic field $H_{c}\left( T\right) $ and $H_{c2}\left( T\right) $
that takes a role of the supercooling field. In strongly layered WSM
superconductors the mean field GL approach is not sufficient due to thermal
fluctuations despite relatively low critical temperatures.

Strength of thermal fluctuations is estimated generally and its is found
that they are strong enough in strongly layered materials to cause Abrikosov
vortex melting. Moreover we predict that, while for type I WSM the
fluctuations of the layered material in magnetic field are three
dimensional, they become two dimensional in the type II phase. Results are
well fitted (see Fig. 4) by general melting line formulas derived within the
lowest Landau level GL approach.

Main results of the paper are applied to the layered WSM superconductor $%
MoTe_{2}$. Magnetic properties of this material were extensively studied\cite%
{MoTe2melting} under pressures from ambient to $30GPa$. In this system the
superconducting critical temperature has maximum at the pressure about $%
12GPa $. While the theory naively predicts \cite{Zyuzin, Rosenstein17} sharp
rise of $T_{c}$ at the topological transition between Type I and Type II
phases of WSM, the region of maximum is beyond the range of its validity
(see Fig.1, with dashed red lines indicating the range). We believe however
that two values of pressure at which magnetic properties were
comprehensively measured belong to different phases of WSM. Non-linear shape
of the transition line to the normal state at temperatures below $T_{c}$ ,
see Fig.4, might be explained either by strong fluctuations in the vortex
matter of the second kind superconductor or by spatial inhomogeneity on the
mesoscopic scale. We argue that the first option is more likely, since the
line clearly has a power dependence on temperature near $T_{c}$.

Our results support a view expressed in ref. \cite{MoTe2melting} that
magnetic properties of this dichalcogenides are reminiscent of those of \
the well studied "conventional" layered superconductor $NbSe_{2}$ (perhaps
this is related to the fact that the later also possesses a pronounced multi
- band electronic structure). It is expected that similar materials exhibit
phenomena described theoretically here. In particular it was observed\ very
recently \cite{PdTe2TypeI} that in a dichalcogenides $PdTe_{2}$ $T_{c}$%
decreases slowly with pressure. In this material the pair of type-II Dirac
points disappears at $6.1$ $GPa$, while a new pair of type-I Dirac points
emerges at $4.7$ $GPa$. Therefore the theoretical analysis of this material
is complicated by the fact that for $4.7-6.1$ $GPa$, the type-II and type-I
Dirac cones coexist \cite{PdTe2}. The superconductor $PdTe_{2}$ was recently
classified as a Type II Dirac semimetal with magnetic measurements confirmed
that $PdTe_{2}$ was a first kind superconductor with $T_{c}=1.64$ $K$ and
the thermodynamic critical field of $H_{c}(0)$ $=13.6$ $mT$ (intermediate
state under magnetic field is typical to a first kind superconductor, as
demonstrated by the differential paramagnetic effect \cite{PdTe2TypeI}).
\bigskip This feature is consistent with the magnetic phase diagram of the
present paper, where the first kind superconductivity is predicted in the
Type-II phase of the WSM (see Fig. 4).

The calculation was limited to strongly layered case. The usage of continuum
3D model instead of fully layered Lawrence - Doniach\cite{LD} model is
justified in the present case while. The calculation can be extended to
arbitrary tunneling strength and is in progress.

\bigskip

\bigskip \textit{Acknowledgements.}

We are grateful to T. Maniv, W. B. Jian, N.L. Wang for valuable discussions.
B.R. was supported by NSC of R.O.C. Grants No. 103-2112-M-009-014-MY3 and is
grateful to School of Physics of Peking University and Bar Ilan Center for
Superconductivity for hospitality. The work of D.L. also is supported by
National Natural Science Foundation of China (No. 11274018 and No. 11674007).

\appendix

\section{Gorkov equations in integral form}

\bigskip Gorkov equations Eq.(\ref{FGE}) can be presented in an integral
form:

\begin{equation}
g_{\epsilon \kappa }\left( \mathbf{r,r}^{\prime }\ \omega \right) =\mathrm{g}%
_{\epsilon \kappa }^{1}\left( \mathbf{r}-\mathbf{r}^{\prime },\omega \right)
-\int \mathrm{g}_{\epsilon \theta }^{1}\left( \mathbf{r}-\mathbf{r}^{\prime
\prime },\omega \right) \Delta _{\theta \phi }^{\ast }\left( \mathbf{r}%
^{\prime \prime }\right) f_{\phi \kappa }^{+}\left( \mathbf{r}^{\prime
\prime }\mathbf{,r}^{\prime },\omega \right) ;  \label{A1}
\end{equation}%
\begin{equation}
f_{\beta \kappa }^{+}\left( \mathbf{r,r}^{\prime },\omega \right) =\int
\mathrm{g}_{\beta \alpha }^{2}\left( \mathbf{r-r}^{\prime \prime \prime
},-\omega \right) \Delta _{\alpha \epsilon }^{\ast }\left( \mathbf{r}%
^{\prime \prime \prime }\right) \left[ \mathrm{g}_{\epsilon \kappa
}^{1}\left( \mathbf{r}^{\prime \prime \prime }-\mathbf{r}^{\prime },\omega
\right) -\int \mathrm{g}_{\epsilon \theta }^{1}\left( \mathbf{r}^{\prime
\prime }-\mathbf{r}^{\prime \prime \prime },\omega \right) \Delta _{\theta
\phi }^{\ast }\left( \mathbf{r}^{\prime \prime }\right) f_{\phi \kappa
}^{+}\left( \mathbf{r}^{\prime \prime }\mathbf{,r}^{\prime },\omega \right) %
\right]  \label{A2}
\end{equation}

\bigskip \ Expanding in small order parameter $\Delta ,$ one obtains Eq.\ref%
{delta} :

\begin{eqnarray}
\Delta \left( \mathbf{r}\right) &=&\frac{g^{2}T}{2}\sum \limits_{\omega
}\int \left[
\begin{array}{c}
\left[ \mathrm{g}_{21}^{2}\left( \mathbf{r-r}^{\prime \prime \prime }\right)
\mathrm{g}_{21}^{1}\left( \mathbf{r}^{\prime \prime \prime }-\mathbf{r}%
\right) \right] \sigma _{12}^{x}\sigma _{12}^{x}+\left[ \mathrm{g}%
_{11}^{2}\left( \mathbf{r-r}^{\prime \prime \prime }\right) \mathrm{g}%
_{22}^{1}\left( \mathbf{r}^{\prime \prime \prime }-\mathbf{r}\right) \right]
\sigma _{21}^{x}\sigma _{12}^{x}+ \\
\left[ \mathrm{g}_{12}^{2}\left( \mathbf{r-r}^{\prime \prime \prime }\right)
\mathrm{g}_{12}^{1}\left( \mathbf{r}^{\prime \prime \prime }-\mathbf{r}%
\right) \right] \sigma _{21}^{x}\sigma _{21}^{x}+\left[ \mathrm{g}%
_{22}^{2}\left( \mathbf{r-r}^{\prime \prime \prime }\right) \mathrm{g}%
_{11}^{1}\left( \mathbf{r}^{\prime \prime \prime }-\mathbf{r}\right) \right]
\sigma _{12}^{x}\sigma _{21}^{x}%
\end{array}%
\right] \Delta \left( \mathbf{r}^{\prime \prime \prime }\right)  \label{A3}
\\
&&-\int \mathrm{g}_{\beta \alpha }^{2}\left( \mathbf{r-r}^{\prime \prime
\prime }\right) \mathrm{g}_{\epsilon \theta }^{1}\left( \mathbf{r}^{\prime
\prime }-\mathbf{r}^{\prime \prime \prime }\right) \mathrm{g}_{\phi \zeta
}^{2}\left( \mathbf{r}^{\prime \prime }\mathbf{-r}_{3}\right) \mathrm{g}%
_{\epsilon \kappa }^{1}\left( \mathbf{r}_{3}-\mathbf{r}\right) \Delta
_{\theta \phi }^{\ast }\left( \mathbf{r}^{\prime \prime }\right) \Delta
_{\alpha \epsilon }^{\ast }\left( \mathbf{r}^{\prime \prime \prime }\right)
\Delta _{\zeta \epsilon }^{\ast }\left( \mathbf{r}_{3}\right)  \notag
\end{eqnarray}

\section{ Calculation of the normal GF}

Normal Green function obeyed the equations \ref{Eq.12},\ref{Eq.14}. First
four GF are calculated from the equation

\bigskip
\begin{equation}
L_{\gamma \beta }^{1}\mathrm{g}_{\beta \kappa }^{1}\left( \mathbf{r-r}%
^{\prime }\right) =\delta ^{\gamma \kappa }\delta \left( \mathbf{r-r}%
^{\prime }\right) ,  \label{B1}
\end{equation}%
where $L_{\gamma \beta }^{1}=\left[ \left( i\omega +\mu +i\mathbf{w\nabla }%
_{r}\right) \delta _{\gamma \beta }+\left( -iv\sigma _{\gamma \beta }^{i}\
\nabla _{r}^{i}\right) \right] $ by performing Fourier transform for
different pseudo-spin indexes. In particular for $\gamma =1,\kappa =1\ $it
reads in momentum representation\bigskip \
\begin{eqnarray}
\ \left( i\omega +\mu -\mathbf{wp}\right) g_{11}^{1}\left( \mathbf{p}\right)
+v\left( \ \mathbf{p}^{x}-i\ \mathbf{p}^{y}\right) g_{21}^{1}\left( \mathbf{p%
}\right) &=&1;  \label{B2} \\
\left( i\omega +\mu -\mathbf{wp}\right) g_{11}^{1}\left( \mathbf{p}\right)
+vp\left( \ \cos \varphi -i\sin \varphi \right) g_{21}^{1}\left( \mathbf{p}%
\right) &=&1.  \notag
\end{eqnarray}%
The rest of the normal GF may be obtained by the same method. The second
group of the normal Green functions obey the equations $L_{\gamma \beta
}^{2}g_{0\beta \kappa }^{2}\left( \mathbf{r-r}^{\prime }\right) =\delta
^{\gamma \kappa }\delta \left( \mathbf{r-r}^{\prime }\right) \ $with $%
L_{\gamma \beta }^{2}$ defined in Eq.(\ref{Eq.12}) are obtained by the same
method. \bigskip The GF obtained after solution of these equations are:

\begin{eqnarray}
g_{22}^{1}\left( \mathbf{p}\right) &=&z^{\ast -1}\left( i\omega +\mu -%
\mathbf{wp}\right) ;\text{ \ }g_{12}^{1}\left( \mathbf{p}\right) =-z^{\ast
-1}vpe^{-i\varphi }  \label{B3} \\
g_{11}^{1}\left( \mathbf{p}\right) &=&z^{\ast -1}\left( i\omega +\mu -\
\mathbf{wp}\right) ;\text{ }g_{21}^{1}\left( \mathbf{p}\right) =-z^{\ast
-1}vpe^{i\varphi }  \notag \\
g_{11}^{2}\left( \mathbf{p}\right) \ &=&z^{-1}\ \left( -i\omega +\mu -%
\mathbf{wp}\right) ;\text{ }g_{12}^{2}\left( \mathbf{p}\right)
=-z^{-1}vpe^{i\varphi }  \notag \\
g_{22}^{2}\left( \mathbf{p}\right) &=&z^{-1}\left( -i\omega +\mu -\ \mathbf{%
wp}\right) ;\text{ \ }g_{21}^{2}\left( \mathbf{p}\right)
=-z^{-1}vpe^{-i\varphi }\ ;  \notag \\
z &=&\left( -i\omega +\mu -\mathbf{wp}\right) ^{2}-\left( vp\right) ^{2},
\notag
\end{eqnarray}%
where $\mathbf{p}$ is the 2D momentum and $\varphi $ is the azimuthal angle
in the $p_{x},p_{y}$ plane.

\section{The critical temperature and the linear term in GL expansion.}

\subsection{The critical temperature for 2D case.}

The linear terms in the GL expansion read:

\begin{equation}
a\left( T\right) =T\sum \nolimits_{\omega ,\mathbf{p}}a\left( \mathbf{p}%
\right) -\frac{1}{g^{2}},  \label{C1}
\end{equation}%
with

\begin{equation}
a\left( \mathbf{p}\right) =2\ Z^{-1/2}\left( \left( vp\right) ^{2}\ +\ \
\omega ^{2}+\left( \mu -w_{x}p_{x}\right) ^{2}\right) \ \text{.}  \label{C2}
\end{equation}%
Here $Z$ is defined in Eq.(\ref{Z}). Performing the summation over $\omega
_{n}$, one obtains,$\ \ $%
\begin{equation}
a\left( T\right) \ \ =\frac{1}{4\left( 2\pi \right) ^{2}\ }\int_{\theta
=0}^{2\pi }\int_{p}\Theta \left( -\varepsilon +\mu +\Omega \right) \Theta
\left( \varepsilon -\mu +\Omega \right) \left \{ \frac{p\tanh \left[ \frac{%
\left \vert p\left( 1+w\cos \theta \right) -\mu \right \vert }{2T}\right] }{%
\left \vert p\left( 1+w\cos \theta \right) -\mu \right \vert }+\frac{p\tanh %
\left[ \frac{\left \vert p\left( 1+w\cos \theta \right) +\mu \right \vert }{%
2T}\right] }{\left \vert p\left( 1+w\cos \theta \right) +\mu \right \vert }%
\right \} -\frac{1}{g^{2}}\text{.}  \label{C3}
\end{equation}%
Introducing new variables:%
\begin{equation}
\varepsilon \left( p,\theta \right) =vp+wp_{x}=p\left( 1+w\cos \theta
\right) ;E=vp\left( 1+w\cos \theta \right) -\mu ,  \label{C4}
\end{equation}%
one obtains
\begin{equation}
a\left( T\right) =\frac{\mu }{8\pi v^{2}}f\left( \kappa \right) \left \{
2\left( \log \frac{\Omega }{2T}\text{ }\tanh \left[ \frac{\Omega }{2T\ }%
\right] -\int_{\varepsilon =0}^{\Omega }d\varepsilon \frac{\log \varepsilon
}{\cosh ^{2}\left[ \frac{\varepsilon }{2T}\right] }\right) +\frac{\Omega }{%
\mu }\right \} -\frac{1}{g^{2}}\text{.}  \label{C5}
\end{equation}
In the adiabatic approximation, $\mu >>\Omega $ it gives for coefficients $%
a\left( T\right) $ and the critical temperature $T_{c}$, Eqs.(\ref{linear},%
\ref{Tc}).

\section{Gradient terms C$_{ik}$ and cubic term}

\bigskip In this Appendix the gradient terms in the GL expansion are
calculated.

\subsection{Diagonal gradient terms for 2D case.}

Gradient terms in the GL expansion has the form of (\ref{Eq. 39}).
Substituting the normal GF from Eq. (\ref{B3}), one obtains after a simple
calculations the diagonal gradient terms. In Cartesian coordinate (with cone
vector $w$ is directed along the $x$ axes)\ the tensor $C_{ki}$ is diagonal
while $C_{xy}$ and $C_{yx}$ are zero due to the reflection symmetry in the $%
y $ direction). The diagonal components are

\begin{equation}
C_{xx}\left( w,\mathbf{p}\right) =\frac{1}{2Z}\left \{
\begin{array}{c}
v^{2}\left( 2p_{x}w_{x}\mu -2p_{x}^{2}w_{x}^{2}-2w_{x}\omega
p_{y}+v^{2}p_{x}^{2}-\omega ^{2}+\left( \mu -w_{x}p_{x}\right)
^{2}-v^{2}p_{y}^{2}\right) ^{2} \\
+v^{2}\left( 2p_{x}w_{x}\mu -2p_{x}^{2}w_{x}^{2}+2p_{y}w_{x}\omega
+2v^{2}p_{x}^{2}-\omega ^{2}+\left( \mu -w_{x}p_{x}\right) ^{2}-\left(
vp\right) ^{2}\right) ^{2} \\
+4v^{2}\left( w_{x}^{2}p_{x}p_{y}-\mu w_{x}p_{y}-p_{y}p_{x}v^{2}-\omega \mu
\right) ^{2}+4v^{2}\left( -w_{x}p_{x}p_{y}w_{x}+p_{y}\mu
w_{x}+p_{y}p_{x}v^{2}-\omega \mu \right) ^{2} \\
+2\left( -\omega ^{2}w_{x}+w_{x}\left( \mu -w_{x}p_{x}\right)
^{2}+2v^{2}p_{x}\left( \mu -w_{x}p_{x}\right) +w_{x}\left( vp\right)
^{2}\right) ^{2}+8\left( \omega w_{x}\left( \mu -w_{x}p_{x}\right)
+v^{2}p_{x}\omega \right) ^{2}%
\end{array}%
\right \} ;  \label{D1}
\end{equation}

\begin{equation}
C_{yy}\left( w,\mathbf{p}\right) =\frac{1}{Z}\left \{
\begin{array}{c}
v^{2}\left( \omega ^{2}-2v^{2}p_{y}^{2}-\left( \mu -w_{x}p_{x}\right)
^{2}+\left( vp\right) ^{2}\right) ^{2} \\
+v^{2}\left[ 4\left( \left( v^{2}p_{y}p_{x}\right) ^{2}+\omega ^{2}\left(
\mu -w_{x}p_{x}\right) ^{2}\right) \right] +4\left( v^{2}p_{y}\right)
^{2}\left( \omega ^{2}+\left( \mu -w_{x}p_{x}\right) ^{2}\right)%
\end{array}%
\right \} \text{.}  \label{D2}
\end{equation}

\subsection{Gradient terms and effective coherent lengths for 2D layer}

After integration over momenta $p$ and the azimuthal angle $\varphi $ in the
second term in equation Eq.(\ref{GL17}) can be performed numerically using
the dimensionless variables

\begin{equation}
E=\kappa \epsilon \cos \varphi +\epsilon ;\epsilon =\frac{E}{\kappa \cos
\varphi +1}=\left( E\psi \right) ;\psi \left( \kappa ,\varphi \right) \ =%
\frac{1}{\left( \kappa \cos \varphi +1\right) };,  \label{D3}
\end{equation}

where%
\begin{equation}
x=-\overline{\mu }+E\epsilon =\frac{vp}{T_{c}},\overline{\mu }=\frac{\mu }{%
T_{c}},\overline{\omega }=\frac{\omega }{T_{c}}  \label{D4}
\end{equation}%
As a result one obtains the gradient terms coefficients which are
proportional to the square of the anisotropic coherence lengths depending on
ratio $\kappa =w/v.$

\begin{eqnarray}
\eta _{y} &=&\frac{1}{2\pi \overline{\mu }\ }\sum \limits_{\omega }\int
\left( x+\overline{\mu }\right) dxd\varphi \cdot sign\left[ \kappa \cos
\varphi +1\right] \left \{ \left( \kappa \cos \varphi +1\right) \left(
\overline{\omega }^{2}+x^{2}\right) \left( \overline{\omega }^{2}+\left(
-x+2\left( x+\overline{\mu }\right) \psi \left( \kappa ,\varphi \right)
\right) \right) \right \} ^{-2}  \label{D5} \\
&&\times \left \{
\begin{array}{c}
\left[ \overline{\omega }^{2}+\left( \overline{\mu }-\kappa \left( x+%
\overline{\mu }\right) \psi \left( \kappa ,\varphi \right) \cos \varphi
\right) ^{2}\right] ^{2}-4\left( x+\overline{\mu }\right) ^{2}\psi
^{2}\left( \kappa ,\varphi \right) \cos 2\varphi \left( \overline{\mu }%
-\left( x+\overline{\mu }\right) \kappa \psi \left( \kappa ,\varphi \right)
\cos \varphi \right) ^{2} \\
+\left( x+\overline{\mu }\right) ^{4}\psi ^{4}\left( \kappa ,\varphi \right)
+2\left( x+\overline{\mu }\right) ^{2}\psi ^{2}\left( \kappa ,\varphi
\right) \overline{\omega }^{2}+2\left( x+\overline{\mu }\right) ^{2}\psi
^{2}\left( \kappa ,\varphi \right) \left( \overline{\mu }-\left( x+\overline{%
\mu }\right) \kappa \psi \left( \kappa ,\varphi \right) \cos \varphi \right)
^{2}%
\end{array}%
\right \}  \notag
\end{eqnarray}%
and

\begin{eqnarray}
\eta _{x} &=&\frac{1}{4\pi \overline{\mu }}\sum \limits_{\omega }\int \left(
x+\overline{\mu }\right) dxd\varphi \frac{sign\left( \left( \kappa \cos
\varphi +1\right) \right) \left( \overline{\omega }^{2}+\left( -x+2\left( x+%
\overline{\mu }\right) \psi \left( \kappa ,\varphi \right) \right)
^{2}\right) ^{-2}}{\left( \kappa \cos \varphi +1\right) ^{2}\left( \overline{%
\omega }^{2}+x^{2}\right) ^{2}}\times  \label{D6} \\
&&\times \left[
\begin{array}{c}
\left(
\begin{array}{c}
2\kappa \ \overline{\mu }\left( x+\overline{\mu }\right) \psi \cos \varphi
-2\left( x+\overline{\mu }\right) ^{2}\psi ^{2}\kappa _{\ }^{2}\cos
^{2}\varphi -2\kappa \ \overline{\omega }\left( x+\overline{\mu }\right)
\psi \sin \varphi \\
+\left( x+\overline{\mu }\right) ^{2}\psi ^{2}\cos 2\varphi -\overline{%
\omega }^{2}+\left( \overline{\mu }-\kappa \left( \left( x+\overline{\mu }%
\right) \psi \right) \cos \varphi \right) ^{2}%
\end{array}%
\right) ^{2}+ \\
\ \left(
\begin{array}{c}
2\kappa \ \overline{\mu }\left( x+\overline{\mu }\right) \psi \cos \varphi
-2\kappa _{\ }^{2}\left( \left( x+\overline{\mu }\right) \psi \right)
^{2}\cos ^{2}\varphi +2\kappa \left( x+\overline{\mu }\right) \psi \overline{%
\omega }\sin \varphi \\
+\left( \left( x+\overline{\mu }\right) \psi \right) ^{2}\cos ^{2}\varphi -%
\overline{\omega }^{2}+\left( \overline{\mu }-\kappa \left( x+\overline{\mu }%
\right) \psi \cos \varphi \right) ^{2}-\left( \left( x+\overline{\mu }%
\right) \psi \right) ^{2}\sin ^{2}\varphi%
\end{array}%
\right) ^{2} \\
+4\ \left( \kappa _{\ }^{2}\left( \left( x+\overline{\mu }\right) \psi
\right) ^{2}\sin \varphi \cos \varphi -\kappa \overline{\mu }\left( \left( x+%
\overline{\mu }\right) \psi \right) \sin \varphi -\left( \left( x+\overline{%
\mu }\right) \psi \right) ^{2}\sin \varphi \cos \varphi -\overline{\omega }%
\overline{\mu }\right) ^{2} \\
+4\ \left( -\kappa ^{2}\left( \left( x+\overline{\mu }\right) \psi \right)
^{2}\sin \varphi \cos \varphi \ +\kappa \overline{\mu }\left( \left( x+%
\overline{\mu }\right) \psi \right) \sin \varphi \ +\left( \left( x+%
\overline{\mu }\right) \psi \right) ^{2}\sin \varphi \cos \varphi -\overline{%
\omega }\overline{\mu }\right) ^{2} \\
+2\ \left( -\overline{\omega }^{2}\kappa \ +\ \kappa \ \left( \overline{\mu }%
-\kappa \left( x+\overline{\mu }\right) \psi \cos \varphi \right)
^{2}+2\left( x+\overline{\mu }\right) \psi \cos \varphi \left( \overline{\mu
}-\kappa \left( x+\overline{\mu }\right) \psi \cos \varphi \right) +\kappa
\left( \left( x+\overline{\mu }\right) \psi \right) ^{2}\right) ^{2} \\
+8\ \left( \kappa \overline{\omega }\ \left( \overline{\mu }-\kappa \left(
\left( x+\overline{\mu }\right) \psi \right) \cos \varphi \right) +\overline{%
\omega }\left( x+\overline{\mu }\right) \psi \cos \varphi \right) ^{2}%
\end{array}%
\right]  \notag
\end{eqnarray}%
The results are presented in Eq.(\ref{CxxCyy}) in the text.

\subsection{Cubic Term in GL Expansion}

Substituting GF into Eq.(\ref{Eq. 44}) in text one obtains:

\begin{equation}
b\left( \mathbf{p}\right) =\frac{2\ \left[ \left( vp\right) ^{2}\ +\omega
^{2}+\left( \mu -wp_{x}\right) ^{2}\right] \left[ \left( vp\right) ^{2}\
+\omega ^{2}+\left( \mu +wp_{x}\right) ^{2}\right] }{\left[ \ \omega
^{2}+\left( \mu -wp_{x}-vp\right) ^{2}\right] \left[ \ \omega ^{2}+\left(
\mu -wp_{x}+vp\right) ^{2}\right] \left[ \omega ^{2}+\left( \mu
+wp_{x}-vp\right) ^{2}\right] \left[ \ \omega ^{2}+\left( \mu
+wp_{x}+vp\right) ^{2}\right] }  \label{D7}
\end{equation}%
and after integration over momentum the result is Eq.(\ref{Eq. 44}) with%
\begin{eqnarray}
&&\ \eta =\sum \limits_{\omega }\int_{\varphi }^{2\pi }\int_{x=0}^{\infty
}\left( x+\overline{\mu }\right) \frac{sign\left[ \kappa \cos \varphi +1%
\right] }{\left( \kappa \cos \varphi +1\right) ^{2}}\times  \label{D8} \\
&&\times \frac{\left \{ \left[ \left( x+\overline{\mu }\right) \psi \right]
^{2}\ +\overline{\omega }^{2}+\left( \overline{\mu }-\kappa \left( x+%
\overline{\mu }\right) \psi \cos \varphi \right) ^{2}\right \} \left \{ %
\left[ \left( x+\overline{\mu }\right) \psi \right] ^{2}\ +\overline{\omega }%
^{2}+\left( \overline{\mu }+\kappa \left( x+\overline{\mu }\right) \psi \cos
\varphi \right) ^{2}\right \} }{\ \left \{ \overline{\omega }^{2}+\left(
\overline{\mu }-\kappa \left( x+\overline{\mu }\right) \psi \cos \varphi
-\left( x+\overline{\mu }\right) \psi \right) ^{2}\right \} \ \left \{
\overline{\omega }^{2}+\left( \overline{\mu }-\kappa \left( x+\overline{\mu }%
\right) \psi \cos \varphi +\left( x+\overline{\mu }\right) \psi \right)
^{2}\right \} }.  \notag
\end{eqnarray}%
This was evaluated numerically.

\subsection{\protect\bigskip Gradient term in direction perpendicular to
layers}

In this case the set of the 3D GF is transformed has the presented in the
form \ref{B3} where $\mu $ is replaced by $\mu -\frac{p_{z}^{2}}{2m_{z}}.$
Substituting the modified 3D GF into Eq.(\ref{Eq. 39}) one obtains

\begin{eqnarray}
C_{zz} &=&\frac{g^{2}Ts}{2\left( 2\pi \hbar \right) ^{3}}\sum
\limits_{\omega }\int_{\mathbf{p}}\frac{p_{z}^{2}}{m_{z}^{2}}\Theta \left(
vp+\frac{p_{z}^{2}}{2m_{z}}-\Omega -\mu +wp_{x}\right) \Theta \left( vp+%
\frac{p_{z}^{2}}{2m_{z}}+\Omega -\mu +wp_{x}\right)  \label{D9} \\
&&\frac{4v^{2}p^{2}\ \left[ \omega ^{2}+\left( \mu -\frac{p_{z}^{2}}{2m_{z}}%
-wp_{x}\right) ^{2}\right] +\left[ \omega ^{2}+\left( \mu -\frac{p_{z}^{2}}{%
2m_{z}}-wp_{x}\right) ^{2}\right] ^{2}+2\left( \mu -\frac{p_{z}^{2}}{2m_{z}}%
-wp_{x}\right) ^{2}\left( vp\right) ^{2}+\left( vp\right) ^{4}}{\left[ \
\omega ^{2}+\left( \mu -\frac{p_{z}^{2}}{2m_{z}}-wp_{x}-vp\right) ^{2}\right]
^{2}\left[ \ \omega ^{2}+\left( \mu -\frac{p_{z}^{2}}{2m_{z}}%
-wp_{x}+vp\right) ^{2}\right] ^{2}}\text{,}  \notag
\end{eqnarray}%
where $\Theta \left( x\right) $ is the theta function restricting the
integration area in the Debye shell at the Fermi energy.

Introducing dimensionless variable by

\begin{equation}
\varepsilon =vp/T;\varepsilon _{z}=\frac{p_{z}^{2}}{2m_{z}T};p_{z}=\sqrt[\ ]{%
2m_{z}T\varepsilon _{z}}\text{,}  \label{D10}
\end{equation}%
the coefficient in Eq.(\ref{Czz}) takes a form:%
\begin{eqnarray}
\eta _{z} &=&\sum \limits_{\omega }\int \sqrt{\varepsilon _{z}}d\varepsilon
_{z}\varepsilon d\varepsilon d\varphi \Theta \left( \varepsilon +\varepsilon
_{z}+\overline{\Omega }-\overline{\mu }\right) \Theta \left( \varepsilon
+\varepsilon _{z}-\overline{\Omega }-\overline{\mu }\right)  \label{D12} \\
&&\frac{4\varepsilon ^{2}\ \left[ \overline{\omega }^{2}+\left( \overline{%
\mu }-\varepsilon _{z}-\kappa \varepsilon \cos \varphi \right) ^{2}\right] +%
\left[ \overline{\omega }^{2}+\left( \overline{\mu }-\varepsilon _{z}-\kappa
\varepsilon \cos \varphi \right) ^{2}\right] ^{2}+2\left( \overline{\mu }%
-\varepsilon _{z}-\kappa \varepsilon \cos \varphi \right) ^{2}\varepsilon
^{2}+\varepsilon ^{4}}{\left \{ \left[ \ \overline{\omega }^{2}+\left(
\overline{\mu }-\varepsilon _{z}-\kappa \varepsilon \cos \varphi
-\varepsilon \right) ^{2}\right] \left[ \ \overline{\omega }^{2}+\left(
\overline{\mu }-\varepsilon _{z}-\kappa \varepsilon \cos \varphi
+\varepsilon \right) ^{2}\right] \right \} ^{2}}\text{.}  \notag
\end{eqnarray}%
This equation was evaluated numerically and results presented in Fig. 2c.

\section{Density of states in WSM.}

In this Appendix we calculate the DOS for the normal electrons described by
the Hamiltonian (\ref{eq1}). Using the dispersion relation for a single
electron,

\begin{equation}
E=\varepsilon +\varepsilon _{z}+\varepsilon \kappa \cos \varphi ,  \label{E1}
\end{equation}%
one obtains for electron density (for two sublattices and two spins)
\begin{equation}
n=\frac{4}{\left( 2\pi \right) ^{3}\hbar ^{3}}\int_{p}\varepsilon
d\varepsilon d\varphi dp_{z}\Theta \left( E\left[ \varepsilon ,p\right] -\mu
\right) \text{,.}  \label{E2}
\end{equation}%
The DOS is
\begin{equation}
\frac{dn}{d\mu }=\frac{4}{\left( 2\pi \right) ^{3}\hbar ^{3}v^{2}}\sqrt{%
\frac{m_{z}}{2}}\int_{p}\varepsilon d\varepsilon d\varphi \frac{d\varepsilon
_{z}}{\sqrt{\varepsilon _{z}}}\delta \left( \mu -\varepsilon -\varepsilon
_{z}-\varepsilon \kappa \cos \varphi \right) ,  \label{E3}
\end{equation}%
where new variables were defined as\bigskip \ $\varepsilon _{z}=\frac{%
p_{z}^{2}}{2m_{z}}$. \bigskip Performing integration over $\varepsilon _{z}$%
, one obtains%
\begin{equation}
\frac{dn}{d\mu }=-\frac{4}{\left( 2\pi \right) ^{3}\hbar ^{3}v^{2}}\sqrt{%
\frac{m_{z}}{2}}\int_{p}\frac{\varepsilon d\varepsilon d\varphi }{\sqrt{\mu
-\varepsilon -\varepsilon \kappa \cos \varphi }}=\frac{\mu ^{3/2}\sqrt{2m_{z}%
}}{3\pi ^{2}\hbar ^{3}v^{2}}f  \label{E4}
\end{equation}%
where the angle integral was calculated in Ref. \cite{Rosenstein17}
resulting in $f$.

\bigskip \newpage

\


\begin{thebibliography}{99}
\bibitem{Weng} H. Weng, X. Dai, and Z. Fang, J. Phys. Cond. Matter. \textbf{%
28}, 303001 (2016). A. Bansil, H. Lin, and T. Das, Rev. Mod. Phys. \textbf{88%
}, 021004 (2016); H. Weng, C. Fang, Z. Fang, B. A. Bernevig, and X. Dai,
Phys. Rev. X\textbf{\ 5}, 011029 (2015); B.Q. Lv, H.M. Weng, B.B. Fu, X.P.
Wang, H. Miao, J. Ma, P. Richard, X.C. Huang, L.X. Zhao, G.F. Chen, Z. Fang,
X. Dai, T. Qian, and H. Ding, Phys. Rev. X \textbf{5}, 031013 (2015); S.-Y.
Xu et al., Science \textbf{349}, 613 (2015).

\bibitem{MoTeearly} L. Huang, T. M. McCormick, M. Ochi, Z. Zhao, M.-T.
Suzuki, R. Arita, Y. Wu, D. Mou, H. Cao, J. Yan, N. Trivedi \& A. Kaminski ,
Nature Materials \textbf{15}, 1155 (2016); Y. Wang et al, Nature Com.
\textbf{7}, 13142 (2016); K. Deng, et al., Nature Physics \textbf{12}, 1105
(2016).

\bibitem{ZrTe} J. Cao, S. Liang, C. Zhang, Y. Liu, J. Huang, Z. Jin, Z.-G.
Che, Z. Wang, Q. Wang, J. Zhao, S. Li, X. Dai, J. Zou, Z. Xia, L. Li and F.
Xiu, Nat. Comm. \textbf{6}, 7779 (2015); W. Yu, Y. Jiang, J. Yang, Z.L. Dun,
H.D. Zhou, Z. Jiang, P. Lu, and W. Pan, Scientific Rep. \textbf{6}, 35357
(2016).

\bibitem{chiral} Y.-Y. Lv, X. Li, Bin-Bin Zhang, W.Y. Deng, Shu-Hua Yao,
Y.B. Chen, Jian Zhou, Shan-Tao Zhang, Ming-Hui Lu, Lei Zhang, M. Tian, L.
Sheng, and Yan-Feng Chen, Phys. Rev. Lett. \textbf{118}, 096603 (2017); M.
Udagawa and E. J. Bergholtz, Phys. Rev. Lett. \textbf{117}, 086401 (2016).

\bibitem{Soluyanov} A. A. Soluyanov, D. Gresch, Z. Wang, Q. Wu, M. Troyer,
X. Dai \& B. A. Bernevig, Nature \textbf{527}, 495 (2015).

\bibitem{Yu} Z.-M. Yu, Y. Yao, and S. A. Yang, Phys. Rev. Lett. \textbf{117}%
, 077202 (2016).

\bibitem{Brien} T. E. O'Brien, M. Diez, and C. W. J. Beenakker, Phys.Rev.
Lett. \textbf{116}, 236401 (2016).

\bibitem{Goerbig1} S. Katayama, A. Kobayashi, Y. Suzumura, J. Phys. Soc.
Japan \textbf{75}, 054705 (2006); M. O. Goerbig, J. -N. Fuchs, G.
Montambaux, F. Pi\'{e}chon, Phys. Rev. B \textbf{78}, 045415 (2008); M.
Hirata et al,Nature Commun. \textbf{7}, 12666 (2016).

\bibitem{toptransition} Y. Zhou, P. Lu, Y. Du, X. Zhu, G. Zhang, R. Zhang,
D. Shao, X. Chen, X. Wang, M. Tian, J. Sun, X. Wan, Z. Yang, W. Yang, Y.
Zhang, and D. Xing£¬ Phys. Rev. Lett., \textbf{117}, 146402 (2016).

\bibitem{Volovik} G.E. Volovik, JETP Lett. \textbf{105,} 519 (2017); Y. Xu,
F. Zhang, and C. Zhang, Phys. Rev. Lett., \textbf{115}, 265304 (2015).

\bibitem{Goerbig2} M. Monteverde, M. O. Goerbig, P. Auban-Senzier, F.
Navarin, H. Henck, C. R. Pasquier, C. M\`{e}zi\'{e}re, and P. Batail, Phys.
Rev B \textbf{87}, 245110 (2013).

\bibitem{Ye} F. Sun and J. Ye , Phys. Rev. B \textbf{96}, 035113 (2017).

\bibitem{Sun} Y. Sun, S.-C. Wu, M. N. Ali, C. Felser, and B. Yan, Phys. Rev.
B \textbf{92}, 161107(R) (2015); J. Ruan, S.-K. Jian, H. Yao, H. Zhang,
S.-C. Zhang \& D. Xing, Nature Com. \textbf{7} 11136 (2016).

\bibitem{HfTe} Y. Qi, W. Shi, P. G. Naumov, N. Kumar, W. Schnelle, O.
Barkalov, C. Shekhar, H. Borrmann, C. Felser, B. Yan, and S. A. Medvedev,
Phys. Rev. B \textbf{94}, 054517 (2016).

\bibitem{YBCO} P. L. Alireza, G. H. Zhang, W. Guo, J. Porras, T. Loew, Y.-T.
Hsu, G. G. Lonzarich, M. Le Tacon, B. Keimer, and Suchitra E. Sebastian,
Phys. Rev. B \textbf{95}, 100505 (2017).

\bibitem{Zyuzin} M. Alidoust, K. Halterman, and A. A. Zyuzin, Phys. Rev. B
\textbf{95}, 155124 (2017).

\bibitem{Rosenstein17} D. Li, B. Rosenstein, B. Ya. Shapiro, and I. Shapiro,
Phys. Rev. B \textbf{95}, 094513 (2017).

\bibitem{DasSarma} S. Das Sarma and Q. Li, Phys. Rev. B \textbf{88},
081404(R) (2013); P.M.R. Brydon, S. Das Sarma , H.-Y. Hui, and J. D. Sau,
Phys. Rev. B \textbf{90}, 184512 (2014); D. Li, B. Rosenstein, B. Ya.
Shapiro, and I. Shapiro, Phys. Rev. B \textbf{90}, 054517 (2014).

\bibitem{FuBerg} L. Fu and E. Berg, Phys. Rev. Lett. \textbf{105}, 097001
(2010).

\bibitem{frontiers} J.-L. Zhang et al. Front. Phys., \textbf{7,} 193 (2012).

\bibitem{Shapiro14} D. Li, B. Rosenstein, B. Ya. Shapiro, and I. Shapiro,
Phys. Rev. B \textbf{90}, 054517 (2014).

\bibitem{Tamai} A. Tamai, Q.\thinspace S. Wu, I. Cucchi, F.\thinspace Y.
Bruno, S. Ricc\`{o}, T.\thinspace K. Kim, M. Hoesch, C. Barreteau, E.
Giannini, C. Besnard, A.\thinspace A. Soluyanov, and F. Baumberger, Phys.
Rev. X \textbf{6}, 031021 (2016).

\bibitem{MoTe2melting} Y. Qi et al., Nat. Comm. \textbf{7,} 11038 (2016).

\bibitem{NbSe2} K. Ghosh, S. Ramakrishnan, A. K. Grover, G. I. Menon, G.
Chandra, T. V. ChandrasekharRao, G. Ravikumar, P. K. Mishra, V. C. Sahni, C. V. Tomy,
G. Balakrishnan, D. Mck Paul, and S. Bhattacharya, Phys. Rev. Lett. \textbf{%
76}, 4600 (1996).

\bibitem{Abrikosov} A. A. Abrikosov, L. P. Gor'kov, I. E. Dzyaloshinskii,
"Quantum field theoretical methods in statistical physics", Pergamon Press,
New York (1965)

\bibitem{Rosenstein18} B. Rosenstein, B. Ya. Shapiro, D. Li, and I. Shapiro,
Phys. Rev. B \textbf{96} 224517 (2017).

\bibitem{Keterson} J.B. Ketterson and S.N. Song, Superconductivity,
Cambridge University Press, 1999.

\bibitem{RMP} B. Rosenstein and D. Li, Rev. Mod. Phys. \textbf{82}, 109
(2010).

\bibitem{Patashinsky} A.Z. Patashinsky, V.L. Pokrovsky. Fluctuation theory
of phase transitions, Pergamon Press, 1979.

\bibitem{Landau} L.D. Landau and E.M. Lifshitz, Statistical Physics, Course
of Theoretical Physics, V 5, p. 478.

\bibitem{shapiro88} B.Ya. Shapiro, JETP Letters \textbf{46} 569 \ (1987)
[ZETP Pis'ma, \textbf{46} \ 451 (1987)]\newpage

\bibitem{Rosenstein02} D. P.Li, B. Rosenstein, \ Phys. Rev B, \textbf{65}
220504 (2002)

\bibitem{PdTe2TypeI} H. Leng, C. Paulsen, Y. K. Huang, and A. de Visser,
Phys. Rev. B \textbf{96}, 220506(R) (2017)

\bibitem{PdTe2} R. C. Xiao, P. L. Gong, Q. S. Wu, W. J. Lu, M. J. Wei, J. Y.
Li, H. Y. Lv, X. Luo, P. Tong, X. B. Zhu, and Y. P. Sun,

Phys. Rev. B \textbf{96}, 075101 (2017).

\bibitem{LD} W. E. Lawrence and S. Doniach, in Proceedings of the Twelfth
Conference on Low Temperature Physics, Kyoto, 1970, edited by E. Kanda
(Keigaku, Tokyo, 1970), p. 361.
\end{thebibliography}
\end{document}